\documentclass[12pt,a4paper]{article}
\usepackage{subcaption}
\usepackage{ragged2e}
\usepackage{mathtools}
\usepackage{physics}
\usepackage{fullpage}
\usepackage{graphicx}
\usepackage{epstopdf}
\usepackage{epsfig}
\usepackage{amssymb}
\newtheorem{definition}{Definition}

\newtheorem{theorem}{Theorem}
\newtheorem{proof}{Proof}
\usepackage{lipsum}
\usepackage{changepage}
\usepackage{amsmath,lipsum}
\usepackage{geometry}
\usepackage{setspace}
\usepackage{todonotes}
\usepackage{kantlipsum}
\newtheorem{thm}{Theorem}[section] 
\newtheorem{lemma}[thm]{Lemma}
\allowdisplaybreaks

\begin{document}
\begin{flushleft}

\begin{center}
\Huge
\textbf{Entanglement Measures and Monogamy} 

\Large

\bigskip \bigskip \bigskip \bigskip \bigskip \bigskip \bigskip \bigskip 

Mathematics 4th Year Dissertation

2020/21

\bigskip \bigskip

\textit{School of Mathematical Sciences}

\textit{University of Nottingham} 

\bigskip \bigskip

Alexey Lopukhin

\bigskip \bigskip

Supervisor: Prof. Gerardo Adesso

\bigskip \bigskip \bigskip \bigskip \bigskip \bigskip \bigskip \bigskip 

\normalsize

\end{center}

\newpage

\tableofcontents 

\newpage

\begin{center}
\LARGE
Abstract.
\end{center} 

\begin{adjustwidth*}{4em}{4em}
\normalsize
\justify
This dissertation focuses on the entanglement quantification. Specifically, we will go over the properties of entanglement that should be satisfied by a "good" entanglement measure. Then we will have a look at some of the propositions of the entanglement measures that have been made over the years. We will talk in greater detail about the entanglement of formation. We will discuss the proposals of the mathematical representations of another property of entanglement, called monogamy. We will introduce some definitions of monogamous entanglement measures that were proposed and compare them. As an original observation of mine (page 28), I will also show that [C. Lancien, S. Di Martino, M. Huber, M. Piani, G. Adesso and A.Winter Phys. Rev. Lett., 117:060501 (2016).] has already developed most of what is necessary to prove that the entanglement of formation and the regularised entropy of entanglement are monogamous entanglement measures in the sense of the definition that was given in [G. Gour and G. Yu, Quantum 2, 81 (2018).]. 
\end{adjustwidth*}

\justify
\begin{center}
\Large
\section{Introduction}
\end{center}

\normalsize
Once considered a mysterious oddity that may be contradicting already established physical laws, and now thought of as one of the most useful exploits for the future quantum technologies. It would be an understatement to say that entanglement is one of the most studied physical phenomena in the history of quantum mechanics. 

The first time being discussed in 1935 [2], the phenomenon did not have its current name and was instead called by Einstein-Podolsky-Rosen (EPR) as "spooky action at a distance". This "spooky action" can be characterized through the following example. Suppose two particles in remote systems $A$ and $B$ are described by a joint state

\begin{equation}
\ket{\phi_1}=\ket{\psi_A}\otimes\ket{\psi_B}=\frac{(\ket{0}+\ket{1})}{\sqrt{2}}\otimes\frac{(\ket{0}+\ket{1})}{\sqrt{2}}
\end{equation}

then it is said to be separable (not entangled) because the state of particle in system $B$ is always unchanged after a quantum measurement being performed on particle in system $A$. While if the joint state is instead 

\begin{equation}
\ket{\phi_2}=\frac{\ket{0}\otimes\ket{0}+\ket{1}\otimes\ket{1}}{\sqrt{2}}
\end{equation} 

then it is called entangled because the state of the particle in system $B$ is supposedly instantaneously changed after a measurement in remote system $A$. 

Exactly this kind of "spooky action at a distance" made many physicists dissatisfied with entanglement because it supposedly allowed a faster than light travel of information, which goes against the laws of relativity. This was precisely the argument that Einstein-Podolsky-Rosen used to contradict the Niels Bohr's theory that a state of a particles is undetermined unless measured. Only thirty years later in 1964 J. S. Bell [3] exhibited that hidden variable theory (predetermined state of particles before measurement) that Einstein was proposing could not be correct. And indeed, later followed the numerous experiments [4] that proved Bell's theory to be true and thus establishing entanglement and quantum theory to be correct. While the entangled particles, regardless of the distance between them, are considered as a single system instead of sending signals to each other as was proposed by EPR. 

After the confirmation that quantum mechanics are governed by very different laws, the confidence to study subjects such as quantum information has increased. And this created greater advancements in bringing us closer to the second quantum revolution- the influx of quantum technologies like quantum computers. And unsurprisingly many of these advancements can only be achieved through the study of the yet poorly understood quantum entanglement. While one of the most examined aspects of entanglement is its quantification. 

Indeed the previously viewed examples $\ket{\psi_1}$ and $\ket{\psi_2}$ only represent respectively the states that are not entangled and those that are maximally entangled. While most states are only entangled to some degree. In the last couple of decades numerous methods of measuring the quantum entanglement have been proposed. However, it will become evident that none of them are perfect. We will examine the pros and cost of each of the mentioned measurements. After we are introduced to the above notions, we will be prepared to talk about the main subject of this dissertation. As the title says, it is the monogamy of entanglement- a physical phenomenon that does not allow limitless sharing of entanglement across many subsystems. We will be reviewing the various methods of mathematical definitions of monogamous entanglement measures. 

\begin{center}
\Large
\section{Prerequisites: basics of quantum information}
\end{center}

\normalsize
But even before the introduction of the entanglement measures, we need to make sure that the reader has the required knowledge of the mathematical operations that are used for describing them [1]. 

\large
\subsection{Hilbert space}

\normalsize
         
          \begin{definition}
          Hilbert space.\newline Suppose $\mathcal{H}$ is a d-dimensional linear space. An inner product on $\mathcal{H}$ is a map $\bra{u}\ket{v}\rightarrow\mathbb{C}$ where $\ket{u}$, $\ket{v}\in\mathcal{H}$. The following is true for all $\ket{\phi}$, $\ket{\phi'}$, $\ket{\psi}$, $\ket{\psi'}$ $\in \mathcal{H}$:
          
          \begin{adjustwidth*}{0em}{2em}
          1) $\bra{\psi}\ket{\psi}\geq0$ and $\bra{\psi}\ket{\psi}=0\Leftrightarrow \ket{\psi}=0$
          
          2) $\bra{\phi}\ket{\psi+\psi'}=\bra{\phi}\ket{\psi}+\bra{\phi}\ket{\psi'}$
          
          3) $\bra{\phi}\ket{\alpha\psi}=\alpha\bra{\phi}\ket{\psi}$
          
          4) $\bra{\phi}\ket{\psi}^*=\bra{\psi}\ket{\phi}$
\end{adjustwidth*}

So such a finite dimensional linear space $\mathcal{H}$ is called a Hilbert space.
          \end{definition}  

The Dirac notation with "ket" symbol $\ket{\,\,\,\,}$ and "bra" symbol $\bra{\,\,\,\,}$ have the following meaning in the space of column vectors $\mathbb{C}^d$. Suppose $\phi$, $\psi$ $\in \mathbb{C}^d$ where

\begin{equation}
\psi=\begin{pmatrix}
           x_{1} \\
           x_{2} \\
           \vdots \\
           x_{d}
         \end{pmatrix} \,\,\,\,\,\,\,\,\,\,\,\, and \,\,\,\,\,\,\,\,\,\,\,\, \phi=\begin{pmatrix}
           y_{1} \\
           y_{2} \\
           \vdots \\
           y_{d}
         \end{pmatrix} 
\end{equation} 

here when adopting Dirac notation we have $\ket{\psi}=\psi$, and $\bra{\phi}=\begin{pmatrix} y_1^* & y_2^* & \ldots & y_d^* \end{pmatrix}$, therefore the inner product of this Hilbert space for these vectors is 

\begin{equation}
\bra{\phi}\ket{\psi}=\sum_{n=1}^{d} y_{n}^*x_{n} 
\end{equation}

\begin{definition}
Standard basis and orthonormal basis (ONB).
\newline
A set $\{\ket{a_1},\ldots,\ket{a_d}\}$ is a basis of Hilbert space  $\mathcal{H}$, if these vectors are linearly independent and they span the whole space, meaning that any $\ket{a}\in\mathcal{H}$ can be written as

\begin{equation}
\ket{a}=\sum_{n=1}^{d}\alpha_n\ket{a_n}
\end{equation}

where $\alpha_n\in\mathbb{C}$.

And it's an orthonormal basis if and only if 

\begin{equation}
\bra{a_i}\ket{a_j}=\delta_{ij}=\begin{cases} 1 & \mbox{if } i=j\\ 0 & \mbox{if } i\ne{j} \end{cases} 
\end{equation}
 
\end{definition}

\large
\subsection{Linear operators on Hilbert spaces (might not need all after adjoint)}

\normalsize
\begin{definition} An operator. \newline
A linear operator $A$ on a Hilbert space $\mathcal{H}$ is a map $A:\mathcal{H}\rightarrow\mathcal{H}$ which satisfies linearity condition

\begin{equation}
A(\alpha\ket{\psi}+\beta\ket{\phi})=\alpha A\ket{\psi}+\beta A\ket{\phi}
\end{equation}

for all $\ket{\psi},\ket{\phi}\in\mathcal{H}$ and $\alpha,\beta\in\mathbb{C}$ 

Any linear operator on a finite dimensional Hilbert space can be written as a matrix. 
\end{definition}

\begin{definition}
Adjoint.\newline Suppose $A$ is an operator acting on a Hilbert space $\mathcal{H}$, then we call $A^{\dagger}$ the adjoint of $A$ which satisfies

\begin{equation}
\bra{\phi}\ket{A^{\dagger}\psi}=\bra{\phi}A^{\dagger}\ket{\psi}=\bra{\psi}A\ket{\phi}^*=\bra{\psi}\ket{A\phi}^*=\bra{A\phi}\ket{\psi}
\end{equation}

for all $\ket{\phi}, \ket{\psi}\in\mathcal{H}$.

Adjoint of an operator is its transpose with conjugated terms inside.

\end{definition}

Properties of adjoint:

\setlength{\parindent}{5ex} 
 1) $(\alpha A)^{\dagger}=\alpha^*A^{\dagger}$ ($\alpha\in\mathbb{C}$) 
 
 2) $(AB)^{\dagger}=B^{\dagger}A^{\dagger}$ 
 
 3)$(A^{\dagger})^{\dagger}=A$
\setlength{\parindent}{0ex} 

\begin{definition}
Normal operators.
\newline
An operator $N$ acting on Hilbert space $\mathcal{H}$ is a normal operator if 

\begin{equation}
NN^{\dagger}=N^{\dagger}N 
\end{equation} 

\end{definition}

\begin{theorem}
Spectral theorem.\newline Suppose $N$ is a normal operator acting on finite dimensional Hilbert space $\mathcal{H}$, then there is an ONB $\{\ket{a_1},\ldots,\ket{a_d}\}\in\mathcal{H}$ and $\lambda_1,...,\lambda_d\in\mathbb{C}$ such that

\begin{equation}
N=\sum_{n=1}^{d}\lambda_n\ket{a_n}\bra{a_n} 
\end{equation} 

We then can note that 

\begin{equation}
N\ket{a_j}=\sum_{i=1}^{d}\lambda_i\ket{a_i}\bra{a_i}\ket{a_j}=\sum_{i=1}^{d}\lambda_i\ket{a_i}\delta_{ij}=\lambda_j\ket{a_j}
\end{equation}

and therefore $\lambda_j$ is an eigenvalue of $N$ with eigenvector $\ket{a_j}$. If some of the eigenvalues are equal then we sum all projections $\ket{a_i}\bra{a_i}$ that have eigenvalue $\lambda$ and we get $P_{\lambda}$. So we can rewrite $N$ as

\begin{equation}
N=\sum_{\lambda\in\sigma(N)}\lambda P_{\lambda}
\end{equation}

where $\sigma(N)$ is a set of all eigenvalues of $N$. So now all projectors that make up the sum of $N$ are orthogonal and unique.
\end{theorem}

\begin{definition}
Functional calculus. \newline If we have a normal operator with spectral decomposition $N=\sum_{\lambda\in\sigma(N)}\lambda P_{\lambda}$ and a function $f:\mathbb{C}\rightarrow\mathbb{C}$ then $f(N)=\sum_{\lambda\in\sigma(N)}f(\lambda) P_{\lambda}$.
\end{definition}

\begin{definition}
Trace.\newline Let $A$ be an operator acting on $\mathcal{H}$ and let $\{\ket{a_1},\ldots,\ket{a_d}\}$ be an ONB. Then the trace of $A$ is

\begin{equation}
Tr(A)=\sum_{i=1}^d A_{ii}=\sum_{i=1}^d \bra{a_i}A\ket{a_i}
\end{equation}

\end{definition}

Trace has some following basic properties:

\setlength{\parindent}{5ex} 
1) $Tr(\alpha A+\beta B)=\alpha Tr(A)+\beta Tr(B)$

2) $Tr(AB)=Tr(BA)$

3) If $\{\ket{a_1},\ldots,\ket{a_d}\}$ and $\{\ket{b_1},\ldots,\ket{b_d}\}$ are two different ONB's then 

\begin{equation}
Tr(A)=\sum_{i=1}^d \bra{a_i}A\ket{a_i}=\sum_{i=1}^d \bra{b_i}A\ket{b_i}
\end{equation}

4) $Tr(\ket{\psi}\bra{\phi})=\bra{\phi}\ket{\psi}$
\setlength{\parindent}{0ex} 

where $A$ and $B$ are operators acting on $\mathcal{H}$ and $\alpha,\beta\in\mathbb{C}$

\large
\subsection{Tensor product of Hilbert spaces}

\normalsize 

Before talking about Hilbert spaces we first look at tensor product of two vectors. If we have two column vectors defined as

\begin{equation}
\ket{\psi}=\begin{pmatrix}
           x_{1} \\
           x_{2} \\
           \vdots \\
           x_{n}
         \end{pmatrix} \,\,\,\,\,\,\,\, and \,\,\,\,\,\,\,\, \ket{\phi}=\begin{pmatrix} 
           y_{1} \\
           y_{2} \\
           \vdots \\
           y_{m}
         \end{pmatrix} \,\,\,\,\,\,\,\, then \,\,\,\,\,\,\,\, \ket{\psi}\otimes\ket{\phi}=\begin{pmatrix}
           x_{1}y_1 \\
           \vdots \\
           x_{1}y_m \\
           \vdots \\
           x_{n}y_1 \\
           \vdots \\
           x_{n}y_m \\
         \end{pmatrix} 
\end{equation}
          
\begin{definition} 
Tensor product of Hilbert spaces.\newline Suppose $\mathcal{H}_1$ and $\mathcal{H}_2$ are two Hilbert spaces, then tensor product $\mathcal{H}_1\otimes\mathcal{H}_2$ is a linear space spanned by $\ket{\phi}\otimes\ket{\psi}=\ket{\phi\otimes\psi}$ where $\ket{\phi}\in\mathcal{H}_1$ and $\ket{\psi}\in\mathcal{H}_2$. We conclude that $\mathcal{H}_1\otimes\mathcal{H}_2$ is a Hilbert space if an inner product on $\mathcal{H}_1\otimes\mathcal{H}_2$ is a map $\bra{\phi\otimes\psi}\ket{\phi'\otimes\psi'}\rightarrow \mathbb{C}$. In particular

\begin{equation}
\bra{\phi\otimes\psi}\ket{\phi'\otimes\psi'}=\bra{\phi}\ket{\phi'}\bra{\psi}\ket{\psi'} 
\end{equation}

where $\ket{\phi},\ket{\phi'}\in\mathcal{H}_1$ and $\ket{\psi},\ket{\psi'}\in\mathcal{H}_2$.

\end{definition} 

\begin{definition} Tensor product of operators.
\newline
Suppose $A:\mathcal{H}_1\rightarrow\mathcal{H}_1$ and $B:\mathcal{H}_2\rightarrow\mathcal{H}_2$ are linear operators, then $A\otimes B$ is a linear operator on $\mathcal{H}_1\otimes\mathcal{H}_2$ that acts in the following way

\begin{equation}
A\otimes B:\ket{\phi_1}\otimes\ket{\phi_2}\rightarrow A\ket{\phi_1}\otimes B\ket{\phi_2} 
\end{equation}
\end{definition}

Suppose 

\begin{equation}
A=\begin{bmatrix}
    A_{11} &    A_{12}                &  \dots     & A_{1d_1}                          &       \\
    A_{21} &   \dots          &  &  &       \\
    \vdots         &                      &       & \vdots                    &       \\
            &                             &       &                    &       \\
      A_{d_11}      &                             & \dots &                          A_{d_1d_1}  & 
    \end{bmatrix} 
    \end{equation}
    
    and $B\in M_{d_2 d_2}$ then

\begin{equation}
A\otimes B=\begin{bmatrix}
    A_{11}B &    A_{12}B                &  \dots     & A_{1d_1}B                           &       \\
    A_{21}B &   \dots          &  &  &       \\
    \vdots         &                      &       & \vdots                    &       \\
            &                             &       &                    &       \\
      A_{d_11}B      &                             & \dots &                          A_{d_1d_1}B  & 
    \end{bmatrix} \in M_{d_1d_2d_1d_2} 
    \end{equation}
   
Properties for $a, b, c\in\mathbb{C}$ and $A, B, C$ operators:

\begin{adjustwidth*}{0em}{2em}

1) $Tr(A\otimes B)=Tr(A)Tr(B)$ \,\,\,\,\,\,\,\,\,\,\,\,\,\,\,\,\,\,\,\,\,\,\,\,\,\,\,\,\,\,\,\,\,\,\, 2) $(A\otimes B)(C\otimes D)=AC\otimes BD$

3)  $(A\otimes B)^{\dagger}=A^{\dagger}\otimes B^{\dagger}$\,\,\,\,\,\,\,\,\,\,\,\,\,\,\,\,\,\,\,\,\,\,\,\,\,\,\,\,\,\,\,\,\,\,\,\,\,\,\,\,\,\,\,\,\,\,\,\,\,\,\,\,\,\, 

\end{adjustwidth*}

\Large
\subsection{Quantum measurements}
\normalsize
\begin{definition}
Projection valued measure(PVM).\newline $\mathcal{H}$ is the state space of our system. A PVM over $\{\lambda_1,...,\lambda_n\}$ is given by a collection of orthogonal projections $\{P_{\lambda_1},...,P_{\lambda_n}\}$ onto subspace of $\mathcal{H}$ with properties:

\setlength{\parindent}{5ex} 
1) $P_{\lambda_i}P_{\lambda_j}=\delta_{ij}P_{\lambda_i}$ 

2) $\sum_{i=1}^{k}P_{\lambda_i}=I$ (Identity) 
\setlength{\parindent}{0ex} 
\end{definition}

A quantum measurement with outcomes $\{\lambda_1,...,\lambda_n\}$ on a system with state space $\mathcal{H}$ is characterised by PVM $\{P_{\lambda_1},...,P_{\lambda_n}\}$.

Probability of getting outcome $\lambda_i$ while having $\ket{\psi}$ as a vector state of the system before the measurement is 

\begin{equation}
p(\lambda_i)=\bra{\psi}P_{\lambda_i}\ket{\psi}=Tr(\ket{\psi}\bra{\psi}P_{\lambda_i}) 
\end{equation}

And the state vector of the system after the measurement if we got outcome $\lambda_i$ is

\begin{equation}
\ket{\psi_{\lambda_i}}=\frac{P_{\lambda_i}\ket{\psi}}{\sqrt{p(\lambda_i)}} 
\end{equation} 

For a bipartite system similarly we have a quantum measurement with outcomes \newline $\{(a_1,b_1), ... , (a_n,b_1), (a_1,b_2), ... , (a_n,b_m)\}$ with Hilbert space $\mathcal{H}_A\otimes\mathcal{H}_B$ is characterised by PVM $\{P_{a_1}\otimes Q_{b_1},..., P_{a_n}\otimes Q_{b_1},P_{a_1}\otimes Q_{b_2},...,P_{a_n}\otimes Q_{b_m}\}$

We return to a single system Hilbert space, where suppose we have a general density matrix $\rho$ (whose definition will be below) instead of a pure state $\psi$, then the residual state after outcome $\lambda_i$ is 

\begin{equation}
\rho_{\lambda_i}=\frac{P_{\lambda_i}\rho P_{\lambda_i}^*}{p(\lambda_i)}
\end{equation}

where $p(\lambda_i)=Tr(P_{\lambda_i}\rho P_{\lambda_i}^*)$ and $P_{\lambda_i}^*=P_{\lambda_i}$ since it is a projection. 

\Large
\subsection{Additional definitions and theorems}

\normalsize

\begin{definition} Partial trace.\newline Suppose we have an operator $A\otimes B$ on $\mathcal{H}_1\otimes\mathcal{H}_2$, then the partical traces are defined by:

\setlength{\parindent}{5ex} 
1) $Tr_1(A\otimes B)=BTr(A)$

2) $Tr_2(A\otimes B)=ATr(B)$
\setlength{\parindent}{0ex} 
\end{definition}

and they follow the same properties as regular traces. 

\begin{definition}
Density matrix.\newline An operator $\rho$ is a density matrix on the system $\mathcal{H}$ if it has the following properties:

\begin{adjustwidth*}{0em}{2em}

1) $\bra{\psi}\rho\ket{\psi}\geq0$ for all $\ket{\psi}\in \mathcal{H}$

2) $Tr(\rho)=1$

\end{adjustwidth*}

Any quantum state is a density matrix. If $\ket{\phi}\in\mathcal{H}_1\otimes\mathcal{H}_2$ then $\rho_1=Tr_2(\ket{\phi}\bra{\phi})$ and $\rho_2=Tr_1(\ket{\phi}\bra{\phi})$ are partial states of systems with $\mathcal{H}_1$ and $\mathcal{H}_2$ Hilbert spaces respectively. 
\end{definition}

\begin{definition}
Mixed and pure states.\newline If suppose $\rho_1=\ket{\psi}\bra{\psi}$ for some $\ket{\psi}\in\mathcal{H}_1$, then we call $\rho_1$ a pure state. Then the state of system with Hilbert space $\mathcal{H}_1$ can be written as $\rho_1=\ket{\psi}$. Otherwise, it's called a mixed state. 
\end{definition}

\begin{theorem}
Schmidt decomposition. \newline Suppose we have a joint pure state $\ket{\psi}$ in a composite system with Hilbert space $\mathcal{H}_A\otimes\mathcal{H}_B$ with $dim(\mathcal{H}_A)=d_A$ and $dim(\mathcal{H}_B)=d_B$. Then there exists $r\leq min(d_A,d_B)$ and ONB's $\{e_1, ... , e_r\}$ in $\mathcal{H}_A$ and $\{e_1, ... , e_r\}$ in $\mathcal{H}_B$ such that 

\begin{equation}
\ket{\psi}=\sum_i^r{\sqrt{\mu_i}\ket{e_i}\otimes\ket{f_i}}
\end{equation}

where $\mu_i>0$ and $\sum_i^r{\mu_i}=1$

\end{theorem}

\begin{center}
\Large
\section{Measuring entanglement}
\end{center}

\normalsize
Now that we made sure we have all the necessary knowledge we can start talking about the entanglement quantification. In this section we will introduce the reader to the various propositions of entanglement measures that have been developed over the years.

\large
\subsection{Entanglement properties}

\normalsize

In order to create a valid measure of entanglement one must first establish properties of entanglement [5,24,44] that must be accounted for. To understand these properties, we must look at the operations that can exploit entanglement. Such operations are called Local Operations and Classical Communication (LOCC) [5]. However, it is very difficult to characterise them as there are so many different possible LOCC's. Therefore, we will instead give one of the most famous examples of LOCC that will give a general understanding of what it represents. The example is the teleportation protocol [21] which works as follows (see Figure 1). 

\begin{center}
\begin{figure}[h]
\centering
\includegraphics[scale=0.42]{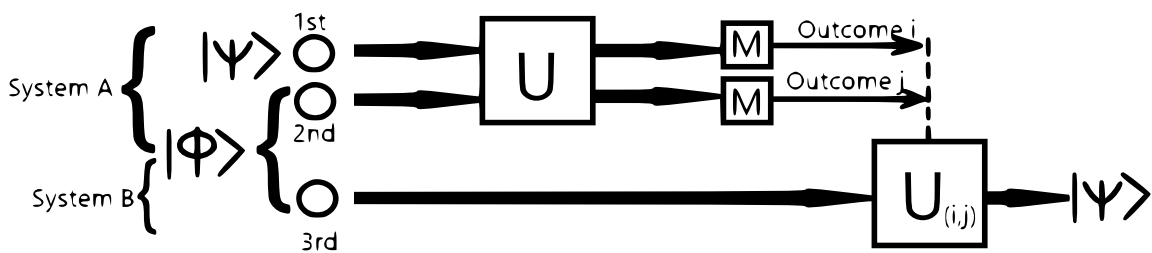}
\caption{The teleportation protocol.}
\end{figure} 
\end{center}

Suppose initially the first and the second particles are in system A and the third particle is in system B. The 1st particle is in state $\ket{\psi}$, the second and third particles make up a joint maximally entangled state $\ket{\phi}$. The joint state of all particles is $\ket{\psi}\otimes\ket{\phi}$. To follow the protocol, we begin by applying the unitary operator $U$ on particles in system A. Then the state of particles in system A is measured. Finally, depending on the outcome $(i,j)$ of the measurement a person in A classically communicates to a person in B to apply unitary $U_{(i,j)}$ on the third particle which results in it having the same state as the initial state of the first particle in A. Thus, through the teleportation protocol a person in A can send state of their particle to a person in B without sending the particle itself. This protocol has even been proven to successfully work at large distances when an experiment has been performed recently in 2017 [6] between ground and a satellite. 

Returning to the initially discussed subject, these are the following properties of entanglement that are to be accounted for when talking about the entanglement measures: 

\begin{adjustwidth*}{2em}{0em}

1) A separable state in a system with Hilbert space $\mathcal{H}_{A_1}\otimes...\otimes\mathcal{H}_{A_n}$ with probability distribution $p_i$ which can be written as

\begin{equation}
\rho_{A_1...A_n}=\sum_i{p_i(\rho^{A_1}_i\otimes...\otimes\rho^{A_n}_i)}
\end{equation}

has no entanglement. If our state $\rho_{A_1...A_n}$ is pure, then if it can be written as a product state $\ket{\psi^{A_1}}\otimes ... \otimes\ket{\psi^{A_n}}$ then it has no entanglement. 

\textit{[See \textbf{Proof 1} in the Appendix]}

2) Entanglement does not increase under LOCC on average. 

So if we have LOCC operation where we have initial state $\rho$ transform into $\rho_i$ with probability $p_i$ then $E(\rho)\geq\sum_i{p_i E(\rho_i)}$, where $E$ is an entanglement measure. The reasoning behind this property has been explained in [24] (Chapter II, page 10). 

3) The maximally entangled state of $\rho_{AB}$ in a composite system with Hilbert space $\mathcal{H}_A\otimes\mathcal{H}_B$ is given by the pure state

\begin{equation}
\frac{\ket{0}\otimes\ket{0}+...+\ket{d-1}\otimes\ket{d-1}}{\sqrt{d}}
\end{equation}

if each partial state of the systems A and B is in d-dimension. To paraphrase, this is the joint state of the particles in systems A and B when there is maximal entanglement between them, given our dimension specification.  

\textit{[See \textbf{Proof 2} in the Appendix]}

4) Additivity: $E(\rho^{\otimes n})=n E(\rho)$ for all $n\in \mathbb{N}$.

So, if between system A and B there are shared $n$ copies of state $\rho$, then the entanglement measured must be $n$ times bigger than the entanglement measure of a single pair.

Though in [24] actually a stronger condition is proposed as a requirement: $E(\rho\otimes\sigma)=E(\rho)+E(\sigma)$ for all $\rho$ and $\sigma$. Though it is often considered too be too demanding, therefore, $E(\rho^{\otimes n})=n E(\rho)$ stands as the property every entanglement measure should satisfy. However, sometimes even this one is considered to be too demanding. 

5) Convexity: $E(\sum_i{p_i\rho_i}) \leq \sum_i{p_i}E(\rho_i)$. 

This property describes the process of the loss of information: going from the individual states $\rho_i$ that have the probability $p_i$ of being prepared to the mixture of these states $\sum_i{p_i\rho_i}$. 

6) Continuity: $E(\rho)-E(\sigma)\rightarrow 0$ as $\abs{\abs{\rho-\sigma}}\rightarrow 0$.

We would want our entanglement measure to be continuous, because we wouldn't want the value of entanglement measure $E(\rho)$ change dramatically when we change $\rho$ only slightly. 

\end{adjustwidth*}

\subsection{Entanglement measure of pure states}

The universally accepted entanglement measure of pure states is the Von Neumann entropy of entanglement, which was introduced in [20] and then was further proven to be a good measure in [7]. Suppose we want to measure entanglement between the particle(s) in system A and the particle(s) in system B that share a joint pure state $\ket{\psi}$ of the composite system with Hilbert space $\mathcal{H}_A\otimes\mathcal{H}_B$, then the Von Neumann entropy of entanglement in this case is given by

\begin{equation}
S(\ket{\psi})=-Tr(\rho_A \log_2(\rho_A))=-Tr(\rho_B \log_2(\rho_B))
\end{equation}

where $\rho_A=Tr_B(\ket{\psi}\bra{\psi})$ and $\rho_B=Tr_A(\ket{\psi}\bra{\psi})$. This is because this entanglement measure perfectly satisfies all of the properties that we introduced.

\textit{[See \textbf{Proof 3} in the Appendix]}

Though the Von Neumann entropy is not exclusively described as just a function that obeys these properties, it also has a physical interpretation [17]. The interpretation goes as follows: suppose we have system A and B that don't share any entanglement, if we want to create a joint state $\ket{\psi}^{\otimes n}$ (meaning that want n pairs of identical state particles between A and B) of the composite system made of A and B, then the number of qubits that must be passed between A and B is equal to $S(\ket{\psi})$.

\large
\subsection{Entanglement measures of mixed states}

\normalsize

However, in a laboratory we would mostly be dealing with mixed states, not the pure states. Therefore, one would be most interested in utilising a method for measuring entanglement of mixed states. Fortunately, there have been made numerous propositions of measures of entanglement of mixed states. However, we will later see that this overabundance of quantity does not necessarily give us the highest quality. The following most renowned propositions are set to measure the entanglement between the particle(s) in system A and the particle(s) in system B that share a joint state $\rho$ of the composite system with Hilbert space $\mathcal{H}_A\otimes\mathcal{H}_B$:

\begin{adjustwidth*}{2em}{0em}
1) Entanglement of formation [17].

\begin{equation}
E_F(\rho)=\min_{\{p_i,\ket{\phi_i}\}}\left(p_i\sum_i{S(\ket{\phi_i})}\right)
\end{equation}

where $\rho=\sum_i{p_i\ket{\phi}}$.

2) Relative entropy of entanglement [9].

\begin{equation}
E_R(\rho)=\inf_{\sigma}\{Tr(\rho log_2(\rho)-\rho log_2(\sigma))\}
\end{equation}

where $\sigma$ stands for all separable states. 

3) Squashed entanglement [10].

\begin{equation}
E_S(\rho=\rho_{AB})=\inf\{I(\rho_{ABE}/2 : Tr_E(\rho_{ABE})=\rho_{AB}\}
\end{equation}

where $I(\rho_{ABE})=S(\rho_{AE})+S(\rho_{BE})-S(\rho_{ABE})-S(\rho_{E})$.

4) Entanglement cost [11].

\begin{equation}
E_C(\rho)=\inf\{r : \lim_{n\rightarrow \infty}(\inf_{\psi}(Tr\abs{\rho^{\otimes n}-\psi(\phi(2^{rn}))}))=0\}
\end{equation}

where $\psi$ stands for a trace preserving LOCC operation and $\phi(K)$ is a maximally entangled state in $K$ dimensions. 

5) Distillable Entanglement [17].

\begin{equation}
E_D(\rho)=\sup\{r : \lim_{n\rightarrow \infty}(\inf_{\psi}(Tr\abs{\psi(\rho^{\otimes n})-\phi(2^{rn})}))=0\}
\end{equation}

where we have the same symbol meaning as in entanglement cost. 

\end{adjustwidth*}

Countless hours of research and problem solving time has been borrowed from numerous scientists and mathematicians regarding the measures of entanglement of mixed states, but unfortunately, to this day we still don't know of the perfect entanglement measure. Indeed, each of the above proposals are known to have flaws.

\begin{adjustwidth*}{0em}{2em}

$\bullet$ Relative entropy of entanglement has been proven to be non-additive [12] (chapter V, subsection B). 

$\bullet$ Squashed entanglement, whose name originates from the fact that $E_D<E_S<E_F$ (proven in [10]), actually has been proven to satisfy all of the entanglement properties: continuity (proven in [13]), additivity, convexity, vanishes for separable states, non-increasing under LOCC (all four proven in [10]). However, there has not been found an easy way to determine the value of squashed entanglement for an arbitrary state.

$\bullet$ Entanglement cost, just like the squashed entanglement, also perfectly satisfies all of the properties because it has been proven in [11] that $E_C(\rho)=\lim_{n\rightarrow \infty}(E_F(\rho^{\otimes n})/n)$, which will be understood once we will go over the properties that entanglement of formation $E_F$ satisfies in the next subchapter. However, again just like the squashed entanglement, it is too difficult to compute. 

$\bullet$ Distillable entanglement is not only difficult to compute like the above two entanglements, but there is also evidence that it is neither additive nor convex [14].

$\bullet$ Lastly there is the entanglement of formation, but it is worth more than just a short mention, therefore we will be discussing it thoroughly in the next subsection. 

\end{adjustwidth*}

\large
\subsection{Entanglement of formation}

\normalsize
Entanglement of formation is the most well-known proposal of entanglement measure of mixed states. There have been put enormous efforts by great number of people into its development. It used to give everyone high hopes of being the ideal entanglement measure until the time when, unfortunately, it was proven to be non-additive [15]. Therefore, the history of the work that has been done on the entanglement of formation should be acknowledged with its own subsection. Also, it makes sense to talk about entanglement of formation in greater detail because we will use some elements from this subchapter later, when we will be talking about monogamy of entanglement. 

So as usual we start with the inspection of whether our entanglement measure satisfies our entanglement properties. And indeed, as mentioned before, it has been demonstrated to satisfy all of them except for additivity. 

\textit{[See \textbf{Proof 4} in the Appendix]}

The physical interpretation of entanglement of formation remains unresolved. It would have had the same interpretation as the Von Neumann entropy of entanglement if it turned out to be additive [19], but as we already know, it has been proven to be non-additive. 

So, we talked about the properties and the physical interpretation of entanglement of formation, but how can we compute it? It is no easy task to derive its value with the formula that we have at the moment. Fortunately, over the course of mainly three papers [17-19] the entanglement of formation of an arbitrary state $\rho$, describing entanglement between two qubits (entanglement between the particle in system A and the particle in system B, each being in their respective two-dimensional partial states), has been proven to be equal to 

\begin{equation}
E_F(\rho)=-\frac{1+\sqrt{1-C^2}}{2}\log_2\left(\frac{1+\sqrt{1-C^2}}{2}\right)-\frac{1-\sqrt{1-C^2}}{2}\log_2\left(\frac{1-\sqrt{1-C^2}}{2}\right)
\end{equation}

where $C=\max\{0, \lambda_1-\lambda_2-\lambda_3-\lambda_4\}$ (concurrence) with $\lambda_i$ being the eigenvalues of $R=\sqrt{\rho\tilde{\rho}}$ in descending order as $i$ increases, where $\tilde{\rho}=(\sigma_y\otimes\sigma_y)\rho^*(\sigma_y\otimes\sigma_y)$ with $\sigma_y$ being the y-Pauli matrix. 

\textit{[See \textbf{Proof 5} (only a summary) in the Appendix]}

Thus, the value of the entanglement of formation is directly computable in the case of two qubits and it satisfies almost all of the entanglement properties. However, as discussed earlier, its flaw is that it is non-additive. It is possible to counter such flaw by introducing a regularised entanglement of formation $E_F^{\infty}(\rho)=\lim_{n\rightarrow \infty}(E_F(\rho^{\otimes n})/n)$ which automatically satisfies additivity (it is simply enough to see what $nE_F^{\infty}(\rho)$ equals to to see additivity). But unfortunately, we do not possess a reliable computation method for the value of the regularised entanglement of formation of an arbitrary mixed state yet. 

So far, we have made a short yet insightful overview of the work that has been done regarding entanglement measures that will also be necessary for our understanding of the next chapter's dedicated topic. 

\begin{center}
\Large
\section{Monogamy of entanglement}
\end{center}

\normalsize

As it was mentioned in the introduction, monogamy of entanglement [22] can be best described as a physical phenomenon that does not allow unlimited distribution of entanglement across many subsystems. It has been mathematically proven to be a valid property of entanglement in [25]. Specifically, the paper shows that for a bipartite state $\rho_{AB}$ there exists $\rho_{AB_1...B_n}$ for an arbitrarily large $n\in\mathbb{N}$, such that $\rho_{AB_1}=\rho_{AB_2}=...=\rho_{AB_n}=\rho_{AB}$, if and only if $\rho_{AB}$ is separable. And additionally, given an entangled $\rho_{AB}$ the paper discovers an upper bound of all the possible $n\in\mathbb{N}$ for which there exits $\rho_{AB_1...B_n}$ such that $\rho_{AB_1}=\rho_{AB_2}=...=\rho_{AB_n}=\rho_{AB}$. 

So why have we not put this physical phenomenon in the subchapter 3.1 as the seventh property of the entanglement? The reason for this is that there still does not exist a single universally agreed mathematical assessment of weather an entanglement measure is monogamous or not. And the possible mathematical characterisation of monogamy is exactly what this chapter examines. 

The best example demonstrating monogamy is the following. Suppose we have three particles in a tripartite pure state in a composite system with Hilbert space $\mathcal{H}_A\otimes\mathcal{H}_B\otimes\mathcal{H}_C$. Each particle is in their respective two-dimensional partial state. And suppose that the first and the second particles are maximally entangled, the ones that are respectively in systems A and B. While the third particle is in system C. Then monogamy manifests itself by not allowing any entanglement shared between the third particle and the other two. So, the joint pure state of the tripartite system must look like this 

\begin{equation}
\ket{\psi_{AB}}\otimes\ket{\psi_{C}}
\end{equation}

where $\ket{\psi_{AB}}$ is the maximally entangled joint  pure state of the first and second particles, and $\rho_C$ is the state of the third particle. This is easily checked if we assume that there is some entanglement between the particle in system $C$ and the particles in the joint systems A and B. If this was the case, then if we were to take a partial trace of the system C then we would get a mixed state

\textit{[See \textbf{Proof 6} in the Appendix]}

which leads us to a contradiction since the joint state of the first and the second particle is supposed to be pure. 

However, this is one of the extreme examples. If those the first and the second particles are only entangled to some degree, then some entanglement can be shared with the third one. So, the main question is: how do we describe this mathematically? 

\large
\subsection{Monogamy of concurrence}

\normalsize

One of the first times the notion of monogamy was captured mathematically in [23]. And it goes as follows. 

\begin{theorem}
For any pure joint sate $\ket{\psi}$ of three qubits in a composite system with Hilbert space $\mathcal{H}_A\otimes\mathcal{H}_B\otimes\mathcal{H}_C$ this inequality is true

\begin{equation}
C^2_{AB}(\rho_{AB})+C^2_{AC}(\rho_{AB})\leq C^2_{A(BC)}(\rho_{ABC})
\end{equation}
\end{theorem}

\textit{[See \textbf{Proof 7} in the Appendix]}

In the \textbf{Theorem 3} $C_{AB}$ stands for concurrence (see subsection 3.4 for its definition) corresponding to the entanglement of formation that measures entanglement between the particles in systems A and B, $C_{AC}$ is analogous and $C_{A(BC)}$ corresponds to entanglement between the particles in system A and the composite system of the systems B and C with Hilbert space $\mathcal{H}_A\otimes\mathcal{H}_B$.

To understand the meaning behind this inequality we must first talk about the concurrence itself. First of all, we must mention that squared concurrence has as much right to be used as a measure of entanglement between two qubits in a joint mixed state as the entanglement of formation has. The reason for this stems from the sebsection 3.4, where we expressed the entanglement of formation as

\begin{equation}
E_F(\rho)=-\frac{1+\sqrt{1-C^2}}{2}\log_2\left(\frac{1+\sqrt{1-C^2}}{2}\right)-\frac{1-\sqrt{1-C^2}}{2}\log_2\left(\frac{1-\sqrt{1-C^2}}{2}\right)
\end{equation}

in terms of concurrence $C$. Let us plot the value of the entanglement of formation against the squared concurrence (see Figure 2).

\begin{center}
\begin{figure}[h]
\centering
\includegraphics[scale=0.8]{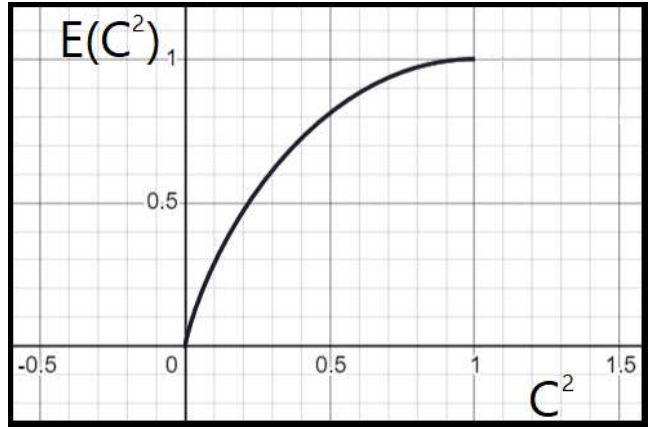}
\caption{The plot of $E_F$ against $C^2$.}
\end{figure} 
\end{center}

As we can see from the plot, the entanglement of formation is a monotonically increasing function of squared concurrence, both are equal to zero for separable states and both are equal to one for maximally entangled states. And it follows all of the other entanglement properties that entanglement of formation follows (like convexity, which has been proven to be satisfied by concurrence in [19]). This confirms our claim about the square concurrence as an entanglement measure of two qubits. 

Now let us return to the inequality (34) to see exactly how it describes monogamy. $C^2_{AB}$ describes the amount of entanglement between the particle in A and the particle in B. $C^2_{AC}$ is analogous. $C^2_{A(BC)}$ describes the amount of entanglement between the particle in A and the two particles in a joint state of composite system with Hilbert space $\mathcal{H}_A\otimes\mathcal{H}_B$, which can be considered as a single particle in a two dimensional state (as seen in Proof 7). So, the inequality (34) means that the amount of entanglement between the particle in A and the other two particles bounds the amount of entanglement between the particles in systems A and B plus the amount of entanglement between the particles in A and C. So now we see how the inequality (34) describes limitations on the amount of entanglement being distributed across a tripartite system, which is a clear demonstration of monogamy. Let us plot $C^2_{AB}$ against $C^2_{AC}$ to visualise this better (see Figure 3).

\begin{center}
\begin{figure}[h]
\centering
\includegraphics[scale=0.5]{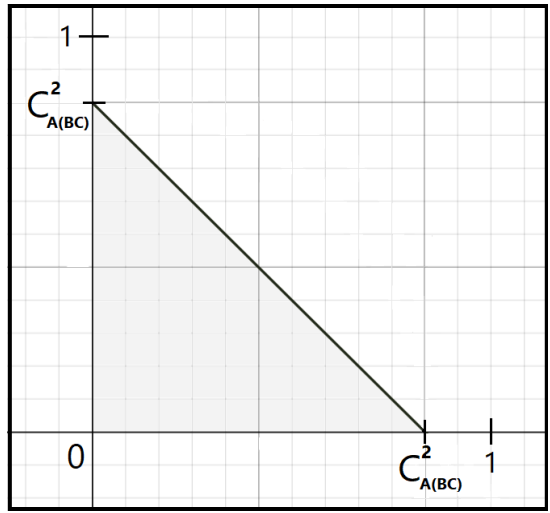}
\caption{The plot of $C^2_{AB}$ against $C^2_{AC}$.}
\end{figure} 
\end{center}

When plotting the inequality (34) one must remember that the values of squared concurrences can be only found between zero and one. Therefore, we end up with a triangular region. If we look at the plot, then we can see that monogamy manifests itself in the following way. The values of ($C^2_{AB}$, $C^2_{AC}$) can never be found outside of the triangle, which is defined by the axis intersection values $C^2_{A(BC)}$, they are constrained inside of it. It is worth noting that $C^2_{A(BC)}$ changes as $\ket{\psi}$ varies. Therefore, the value of $C^2_{A(BC)}$ moves between $0$ and $1$, resulting in the hypotenuse of the triangle going back and forth as $\rho$ changes. Though our visual definition of monogamy still remains to be true, nevertheless. This is because regardless of the size of the triangle we will always be only able to find the value of ($C^2_{AB}$, $C^2_{AC}$) only inside the triangle for each $\ket{\psi}$. 

The inequality (34) does not work for all tripartite mixed states $\rho$. This is because we cannot treat state $\rho_{BC}$ as a two-dimensional state since it does not always have at least two non-zero eigenvalues, which implies that we cannot use the entanglement of formation formula (36) since it only works for measuring the entanglement of a joint state of two qubits. And therefore $C^2_{A(BC)}$ is undefined. 

But we can come up with a similar inequality, which is true for all tripartite mixed states $\rho$ of three qubits in a composite system with Hilbert space $\mathcal{H}_A\otimes\mathcal{H}_B\otimes\mathcal{H}_C$

\begin{equation}
C^2_{AB}+C^2_{AC}\leq \min_{p_i,\ket{\psi_i}}\left(\sum_i p_iC^2_{A(BC)}(\ket{\psi_i}\bra{\psi_i})\right)
\end{equation}

where $\rho=\sum_i p_i\ket{\psi_i}\bra{\psi_i}$. 

\textit{[See \textbf{Proof 8} in the Appendix]}

This inequality also displays monogamous properties of squared concurrence. Just like the inequality (34) for pure states it restricts the possible values of $(C^2_{AB},C^2_{AC})$ for each tripartite state $\rho_{ABC}$. 

So far, we have talked about mathematical expressions showing the monogamy of concurrence only. But what about other entanglement measures? Suppose again we have pure tripartite state of three qubits in a system with Hilbert space $\mathcal{H}_A\otimes\mathcal{H}_B\otimes\mathcal{H}_C$. Then it can be shown with a simple example that inequality

\begin{equation}
E_{AB}+E_{AC}\leq E_{A(BC)}
\end{equation}

does not work for the entanglement of formation. If we have a pure tripartite state 

\begin{equation}
\frac{1}{\sqrt{2}}(\ket{100}+\frac{1}{\sqrt{2}}\ket{010}+\frac{1}{\sqrt{2}}\ket{001})
\end{equation}

then we get $E_{AB}\approx0.6$, $E_{AC}\approx0.6$ and $E_{A(BC)}=1$ which contradicts the inequality (34). But that does not really mean that entanglement of formation does not follow the monogamy property. Indeed, the inequality (34) being true for all tripartite pure states of three qubits is a legitimate mathematical characterisation which shows that concurrence is a monogamous measure of entanglement. The same inequality has even been proven (in [26], chapter III, page 4) to be valid for the squashed entanglement and the one-way distillable entanglement for all dimensions of the tripartite state $\rho$ (the one-way distillable entanglement is the type of distillable entanglement where classical communications can only go one way: from system A to system B [27]). But that does not mean that this must be the test of monogamy for all other entanglement measures as well. Because there are indeed other entanglement measures that do not follow the inequality (34), just like the entanglement of formation or the distillable entanglement. But what if these entanglement measures could follow some different inequality for all tripartite states that could indicate monogamous properties for them? 

\large
\subsection{Monogamy of other entanglement measures}

\normalsize
Following the ending of the previous subchapter we would like to come up with a singular inequality for any entanglement measure which is a satisfying test of weather that entanglement measure is monogamous or not. We will try a similar approach as in the inequality (34), but much more flexible [28]. And we will see weather our new inequality will indicate monogamous properties of the entanglement measures like the entanglement of formation and the relative entropy of entanglement. 

So let us try the following statement as an assessment of weather an entanglement measure $E$ is monogamous. 

\begin{definition} If there exists a function $f:R\times R\rightarrow R$ (where $R$ is the real numbers in $[0,\infty)$ interval) such that 

\begin{equation}
E_{A(BC)}(\rho_{ABC})\geq f(E_{AB}(\rho_{AB}),E_{AC}({\rho_{AC}}))
\end{equation}

is true for all tripartite states $\rho_{ABC}$ in a composite system with Hilbert space $\mathcal{H}_A\otimes\mathcal{H}_B\otimes\mathcal{H}_C$, then the entanglement measure $E$ is monogamous. 
\end{definition}

Here the lower indexes have the same meaning as in the inequality (34). So, for example, $E_{AB}$ stands for an entanglement measure between the particles in systems A and B.

However, right now this statement as a verification of monogamy is much broader than what we should have been aiming for. The reason for this is that just local operations [24] (without classical communication) do not increase entanglement. Therefore, entanglement must not increase under partial trace. So, this means that 

\begin{equation}
E_{A(BC)}(\rho_{ABC})\geq E_{AB}(\rho_{AB}) \,\,\,\,\, and \,\,\,\,\, E_{A(BC)}(\rho_{ABC})\geq E_{AC}(\rho_{AC})
\end{equation}

by default. This means that with our current test of whether an entanglement measure is monogamous we can just pick $f(E_{AB}(\rho_{AB}),E_{AC}({\rho_{AC}}))=\max(E_{AB}(\rho_{AB}),E_{AC}(\rho_{AC}))$ to satisfy the inequality (40) and, therefore, make any entanglement measure monogamous automatically, making our definition useless. Therefore, since

\begin{equation}
E_{A(BC)}(\rho_{ABC})\geq \max(E_{AB}(\rho_{AB}),E_{AC}(\rho_{AC}))
\end{equation}

we should restrict our inequality inside the Definition 14 to instead be

\begin{equation}
E_{A(BC)}(\rho_{ABC})\geq f(E_{AB}(\rho_{AB}),E_{AC}({\rho_{AC}}))>\max(E_{AB}(\rho_{AB}),E_{AC}(\rho_{AC}))
\end{equation}

Though a small correction should be made to this inequality. We should be allowed to have $f=\max(E_{AB},E_{AC})$ for some values of $E_{AB}$ and $E_{AC}$. We just need to make sure there exists a two-dimensional area of $(E_{AB},E_{AC})$ for which we strictly have $f(E_{AB},E_{AC})>\max(E_{AB},E_{AC})$ (this way we still don't get every entanglement measure to be monogamous by default by our new definition). So let us instead have this inequality 

\begin{equation}
E_{A(BC)}(\rho_{ABC})\geq f(E_{AB}(\rho_{AB}),E_{AC}({\rho_{AC}}))\gtrdot\max(E_{AB}(\rho_{AB}),E_{AC}(\rho_{AC}))
\end{equation}

inside the Definition 14 instead. Where the meaning behind $\gtrdot$ is precisely what we explained in the previous paragraph: there can be equality for some values of $E_{AB}$ and $E_{AC}$, but there must be a space of $(E_{AB},E_{AC})$ for which $f(E_{AB},E_{AC})>\max(E_{AB},E_{AC})$. Thus, from now on when we talk about the Definition 14, we mean that it has the inequality (43) instead of the inequality (39). 

\begin{center}
\begin{figure}[h]
\centering
\includegraphics[scale=0.5]{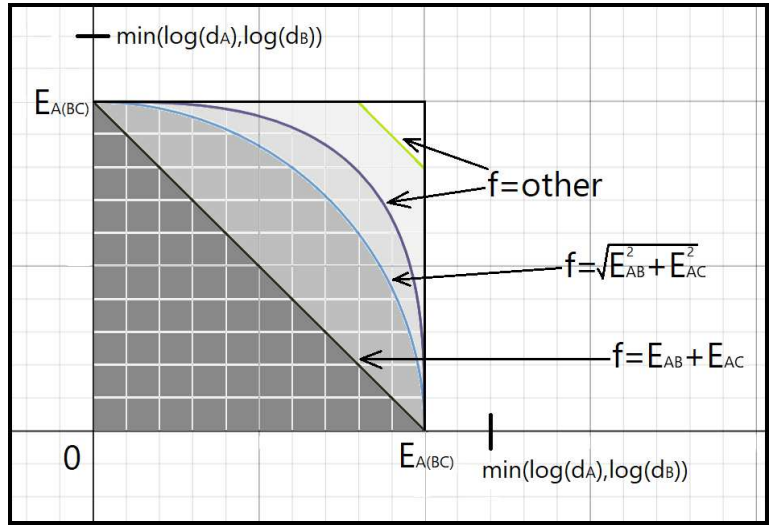}
\caption{The possible plots of $E_{AB}$ against $E_{AC}$.}
\end{figure} 
\end{center}

For the visualisation of the Definition 14 let us give some examples (see Figure 4) of the functions $f(E_{AB},E_{AC})$ that would satisfy the inequality (43). For example, $f=E_{AB}+E_{AC}$ meets the requirements. We have $f=E_{AB}+E_{AC}=\max(E_{AB},E_{AC})$ only for $(0,E_{A(BC)})$ and $(E_{A(BC)},0)$. This is an acceptable correction because for the rest of the space of $(E_{AB},E_{AC})$ we have $f(E_{AB},E_{AC})>\max(E_{AB},E_{AC})$, as required. The function $f=\sqrt{E_{AB}^2+E_{AC}^2}$ also satisfies the inequality in an identical way to the previous one. And then there are other functions. For example, even a function defined like

\begin{equation}
f=\begin{cases}
      \max(E_{AB},E_{AC}) & \text{for}\,\,\, E_{AB},E_{AC}\in[0,\frac{4}{5}E_{A(BC)})\\ 
      E_{AB}+E_{AC}-\frac{4}{5}E_{A(BC)} & \text{for}\,\,\, E_{AB},E_{AC}\in(\frac{4}{5}E_{A(BC)},E_{A(BC)}]
    \end{cases}
\end{equation}

also satisfies the inequality (43). This is because, again, there exists a space of $(E_{AB},E_{AC})$ such that $f(E_{AB},E_{AC})>\max(E_{AB},E_{AC})$, namely the space $E_{AB},E_{AC}\in(\frac{4}{5}E_{A(BC)},E_{A(BC)}]$. For this function, the inequality (43) generates area filling most of the square except for the small corner at the top right, as seen in the Figure 4.

The reason why $E_{A(BC)}\leq \min(\log_2(d_A),\log_2(d_B))$ (as seen in the Figure 4 on the both axes) is shown in both: the end of the Proof 9 and the beginning of the Proof 10. In Figure 4 we can see that $E_{A(BC)}\geq \max(E_{AB},E_{AC})$ is specified by a square region with the length of an edge that equals to $E_{A(BC)}$. 

As promised, now we inspect weather the entanglement of formation and relative entropy of entanglement are monogamous under our new broader definition. It turns out that they do not. 

\textit{[See \textbf{Proof 9} in the Appendix]}

Then we can show that even their regularised versions, $\lim_{n\rightarrow \infty}(E_F(\rho^{\otimes n})/n)$ and \newline$\lim_{n\rightarrow \infty}(E_R(\rho^{\otimes n})/n)$, do not follow the new definition of monogamous entanglement measures either. 

\textit{[See \textbf{Proof 10} in the Appendix]}

These entanglement measures were considered to be perfect, following every entanglement property (remember, the regularised versions of the two entanglement measures follow the additivity property, while the non-regularised don't). But they do not display monogamous properties even in the sense of our very flexible inequality. 

But perhaps our new definition of monogamous entanglement measures is actually a bit too demanding. If we alter it, so that we check if there exists a function $f$ for every fixed dimension of $\mathcal{H}_A\otimes\mathcal{H}_B\otimes\mathcal{H}_C$ and all of the tripartite states $\rho_{ABC}$ of the system with that Hilbert space (we don't consider infinite dimensions), then it can be shown that the entanglement of formation and the regularised relative entropy of entanglement are monogamous in the sense of this newest definition. Specifically, for every tripartite state $\rho_{ABC}$ in a composite system with Hilbert space $\mathcal{H}_A\otimes\mathcal{H}_B\otimes\mathcal{H}_C$ it was found that for the entanglement of formation we have

\begin{equation}
\begin{split}
E^F_{A(BC)}(\rho_{ABC})\geq \max(&E^F_{AB}(\rho_{AB})+\frac{c}{d_Ad_C\log_2(\min(d_A,d_C))^8}E^F_{AC}(\rho_{AC})^8,\\
&E^F_{AC}(\rho_{AC})+\frac{c}{d_Ad_B\log_2(\min(d_A,d_B))^8}E^F_{AB}(\rho_{AB})^8)>\\
&\gtrdot\max(E^F_{AB}(\rho_{AB}),E^F_{AC}(\rho_{AC}))
\end{split}
\end{equation}

and for the regularised relative entropy of entanglement, we have 

\begin{equation}
\begin{split}
E^{R\infty}_{A(BC)}(\rho_{ABC})\geq \max(&E^{R\infty}_{AB}(\rho_{AB})+\frac{c}{d_Ad_C\log_2(\min(d_A,d_C))^4}E^{R\infty}_{AC}(\rho_{AC})^4,\\
&E^{R\infty}_{AC}(\rho_{AC})+\frac{c}{d_Ad_B\log_2(\min(d_A,d_B))^4}E^{R\infty}_{AB}(\rho_{AB})^4)>\\
&\gtrdot\max(E^{R\infty}_{AB}(\rho_{AB}),E^{R\infty}_{AC}(\rho_{AC}))
\end{split}
\end{equation}

where $d_A$, $d_B$ and $d_C$ are the dimensions of the partial states $\rho_A$, $\rho_B$ and $\rho_C$ respectively. 

\textit{[See \textbf{Proof 11} in the Appendix]}

Thus, we see that indeed the values of $(E_{AB},E_{AC})$ are bounded by a function $f$, such that $f\gtrdot\max(E_{AB},E_{AC})$, and therefore, these entanglement measures display monogamous properties. Because we have shown that indeed there is an area of values $(E_{AB},E_{AC})$ that can't exist inside the square given by $E_{A(BC)}\geq\max(E_{AB},E_{AC})$ (remember no value of $E_{AB}$ or $E_{AC}$ could ever be outside of the square by default because of non-increase of entanglement under partial tracing). 

To visualise this, we present plots of $E_{AB}$ against $E_{AC}$ for the respective entanglement measures in the case of a tripartite state of tree qubits in the Figure 5.

\begin{figure}[h]
\centering
\begin{subfigure}{.5\textwidth}
  \centering
  \includegraphics[width=0.8\linewidth]{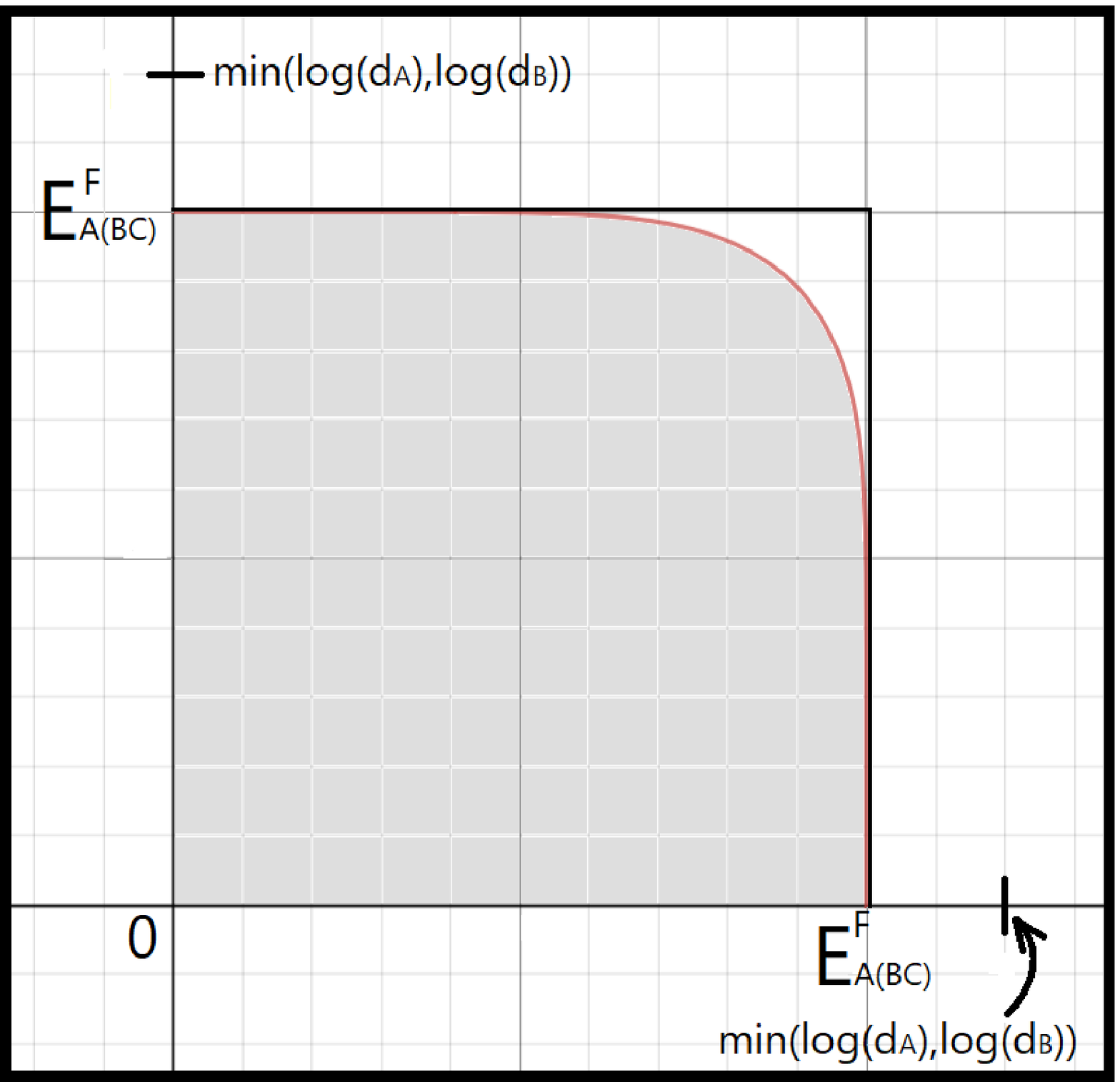}
  \caption{Entanglement of formation.}
  \label{fig:sub1}
\end{subfigure}%
\begin{subfigure}{.5\textwidth}
  \centering
  \includegraphics[width=0.8\linewidth]{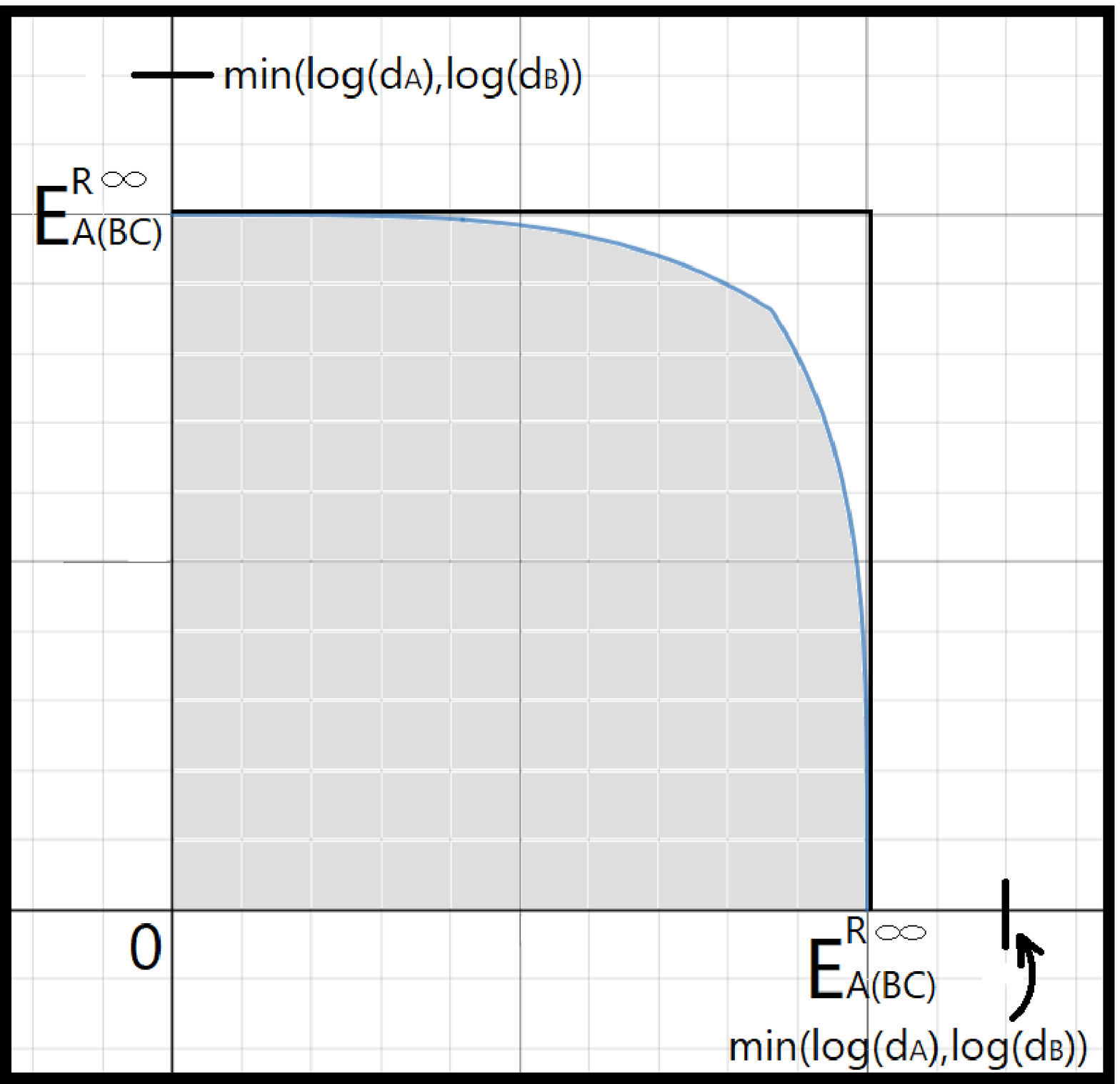}
  \caption{Regularised entropy of entanglement.}
  \label{fig:sub2}
\end{subfigure}
\caption{The two plots of $E_{AB}$ against $E_{AC}$ with $d_A=d_B=d_C=2$.}
\label{fig:test}
\end{figure}

One could compare these plots with the case of the squared concurrence plot in Figure 3, which also dealt with the tripartite states of three qubits. And one has to notice that the inequalities (45) and (46) accomplish the same result as the inequality (34), except that we have shown that the area of the non-existent values of $(E_{AB},E_{AC})$ is smaller for the entanglement of formation and the relative entropy of entanglement. Thus, if the inequality for squared concurrences captures monogamy, then these inequalities do as well. 

However, we must note, as we did already in previous subchapter, that the value of  $E_{A(BC)}^F$ changes as $\rho_{ABC}$ changes. So the area of non-existent values also shrinks or expands as $\rho_{ABC}$ changes. So in order to compare the areas of the non-existent values of $(E_{AB},E_{AC})$  between the three plots, we actually compare these areas relative to the square with the edges of length $E_{A(BC)}$. 

If we increase the dimension of each partial state to $d_A=d_B=d_C=4$ then the area of non-existent values of $(E_{AB},E_{AC})$ will be dramatically decreased, as seen in the Figure 6. 

\begin{center}
\begin{figure}[h]
\centering
\includegraphics[scale=0.4]{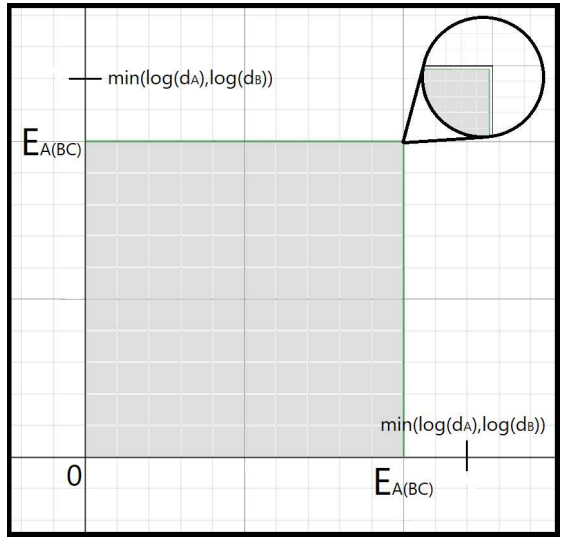}
\caption{The plot of $E_{AB}$ against $E_{AC}$.}
\end{figure} 
\end{center}

And it tends to zero as the the dimensions go to infinity, which was demonstrated in the Proof 9 and the Proof 10. There we showed that for infinite dimensions this area does not exist. 

There is one last thing that we should point out before finishing this subchapter. The entanglement of formation remains to be imperfect even though it was confirmed that it is monogamous. However, now we know that the regularised relative entropy of entanglement follows all of the properties of entanglement. This means that it is in the same group as the squashed entanglement- perfectly satisfying all of the properties, but still very difficult to compute. 

\large
\subsection{Monogamy in terms of equalities?}

\normalsize
The paper [41] argues that the definition of monogamous entanglement measures given in [28] is not the best one. We will give the definition of monogamous measurement that was introduced in [41] and then compare it to the previous definition from [28].

\begin{definition}
An entanglement measure $E$ is monogamous if for all tripartite states $\rho_{ABC}$ of composite system with Hilbert space $\mathcal{H}_A\otimes\mathcal{H}_B\otimes\mathcal{H}_C$ such that

\begin{equation}
E_{A(BC)}(\rho_{ABC})=E_{AB}(\rho_{AB})
\end{equation}

we have $E_{AC}(\rho_{AC})=0$
\end{definition}

So visually the Definition 15 means that an entanglement measure $E$ is monogamous if the values of $(E_{AB}(\rho_{AB}),E_{AC}(\rho_{AC}))$ can only be found on the interior of the square generated by 

\begin{equation}
E_{A(BC)}(\rho_{ABC})\geq\max(E_{AB}(\rho_{AB}),E_{AC}(\rho_{AC}))
\end{equation}

and on the two bottom left edges of that square (see Figure 7). To see why this is true, I remind that the values of $(E_{AB}(\rho_{AB}),E_{AC}(\rho_{AC}))$ can never be found outside of the square generated by (48). Because entanglement is non-increasing under local operations, specifically the partial tracing. 

\begin{center}
\begin{figure}[h]
\centering
\includegraphics[scale=0.45]{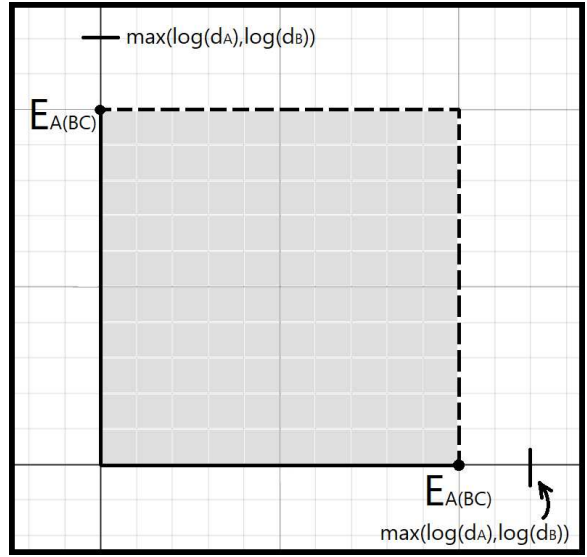}
\caption{The plot of $E_{AB}$ against $E_{AC}$.}
\end{figure} 
\end{center}

It is easy to confirm that with this newest definition of monogamy entanglement measures such as squared concurrence (only for 3 qubits), entanglement of formation and regularised relative entropy of entanglement are monogamous. For the squared concurrence one just needs to insert $C^2_{A(BC)}(\rho_{ABC})=C^2_{AB}(\rho_{AB})$ into the inequality (34) to see that the only possible value of $E_{AC}(\rho_{AC})$ is zero. For the entanglement of formation, the proof is given below. The proof for the regularised relative entropy of entanglement is the same. 

\textit{[See \textbf{Proof 12} in the Appendix]}

But at this moment it looks more intuitive than a solid proof that the inequalities (45) and (46) show that their respective entanglement measures are monogamous in the sense of the Definition 15. The inequalities are dimension dependent while the Definition 15 is not. The detailed proof that the two entanglement measures are monogamous in terms of the Definition 15 will be given under the Definition 17. 

But before that we need to consider a few things. Let us begin with noting that the newest definition of monogamous entanglement measures (Definition 15) does not agree with the definition from [28], which was introduced in the previous subchapter. This is evident by the fact that the Definition 14 (with the reformed inequality (43) replacing the inequality (39), as was done in the previous subchapter) allows  $E_{AC}(\rho_{AC})$ to be non-zero when $E_{A(BC)}(\rho_{ABC})=E_{AB}(\rho_{AB})$. The best example is the function $f$ given by (44), which allows all of the values of $(E_{AB}(\rho_{AB}),E_{AC}(\rho_{AC}))$ inside the square generated by (48), except for the small top right corner of the square (see Figure 4). 

But the Definition 15 of monogamous entanglement measures becomes almost equivalent to the Definition 14, if we revise it as follows. 

\begin{definition} If there exists a function $f:R\times R\rightarrow R$ (where $R$ is the real numbers in $[0,\infty)$ interval) such that 

\begin{equation}
E_{A(BC)}(\rho_{ABC})\geq f(E_{AB}(\rho_{AB}),E_{AC}({\rho_{AC}}))\geq\max(E_{AB}(\rho_{AB}),E_{AC}({\rho_{AC}}))
\end{equation}

is true for all tripartite states $\rho_{ABC}$ in a composite system with Hilbert space $\mathcal{H}_A\otimes\mathcal{H}_B\otimes\mathcal{H}_C$, and such that 

\begin{equation}
f(E_{AB}(\rho_{AB}),E_{AC}({\rho_{AC}}))=\max(E_{AB}(\rho_{AB}),E_{AC}(\rho_{AC}))
\end{equation}

is only true at $(E_{AB}(\rho_{AB}),E_{AC}(\rho_{AC}))=(0,E_{A(BC)}(\rho_{ABC}))$ and $(E_{AB}(\rho_{AB}),E_{AC}(\rho_{AC}))=$ \newline$=(E_{A(BC)}(\rho_{ABC}),0)$, then the entanglement measure $E$ is monogamous. 
\end{definition}

Visually (see Figure 8) this definition means that an entanglement measure $E$ is monogamous if for all $\rho_{ABC}$ the values of $(E_{AB}(\rho_{AB}),E_{AC}(\rho_{AC}))$ can only be found inside the area confined by the two axes $E_{AB}(\rho_{AB})$ and $E_{AC}(\rho_{AC})$, and some curve $E_{A(BC)}(\rho_{ABC})=f(E_{AB}(\rho_{AB}),E_{AC}({\rho_{AC}}))$, such that it joins together the points $(E_{A(BC)}(\rho_{ABC}),0)$ and $(0,E_{A(BC)}(\rho_{ABC}))$, and such that it is entirely inside (except for its two endpoints) the interior of the square area generated by (48). 

\begin{center}
\begin{figure}[h]
\centering
\includegraphics[scale=0.5]{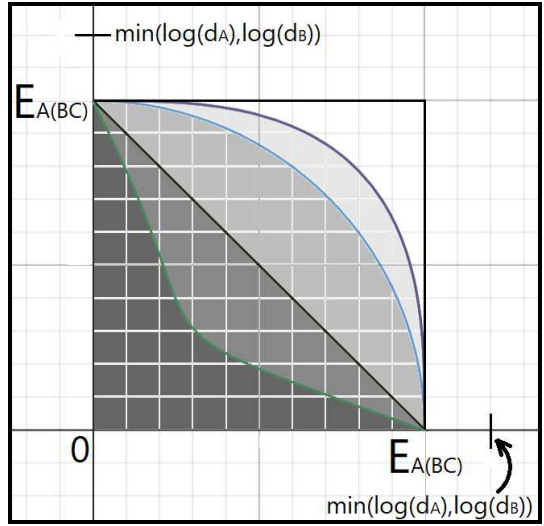}
\caption{The plot of $E_{AB}$ against $E_{AC}$.}
\end{figure} 
\end{center}

Now the Definition 15 and the Definition 16 are indeed almost equivalent. However, at this moment it looks like there remains one difference between them. With the Definition 16 it is impossible to find a curve $E_{A(BC)}(\rho_{ABC})= f(E_{AB}(\rho_{AB}),E_{AC}({\rho_{AC}}))$ such that it engulfs the whole interior of the square generated by (48) while not being the boundary of the said square. Therefore, there can exist entanglement measures that are monogamous in the sense of the Definition 15, but not monogamous in the sense of the Definition 16. While all of the entanglement measures that are monogamous in the sense of the Definition 16, are then automatically monogamous in the sense of the Definition 15. Thus, the definition 16 is the stricter one. 

However, after the following theorem it will finally become clear how to further revise the Definition 16 to make it completely equivalent to the definition 15. 

\begin{theorem}
An entanglement measure $E$ is monogamous in the sense of the Definition 15 if and only if there exists $0<\alpha<\infty$ for every fixed dimension $d=dim(\mathcal{H}_A\otimes\mathcal{H}_B\otimes\mathcal{H}_C)$ such that 

\begin{equation}
E_{A(BC)}(\rho_{ABC})\geq (E_{AB}(\rho_{AB})^{\alpha}+E_{AC}(\rho_{AC})^{\alpha})^{1/\alpha}
\end{equation}

for all tripartite states $\rho_{ABC}$ of a composite system with Hilbert space $\mathcal{H}_A\otimes\mathcal{H}_B\otimes\mathcal{H}_C$. 
\end{theorem}

\textit{[See \textbf{Proof 13} in the Appendix]}

The inequality (51) is one of the earlier representations of monogamy. This inequality was talked about and applied to various entanglement measures in [45]. We can see what the inequality (51) looks like for various $\alpha$'s in the Figure 9. 

\begin{center}
\begin{figure}[h]
\centering
\includegraphics[scale=0.35]{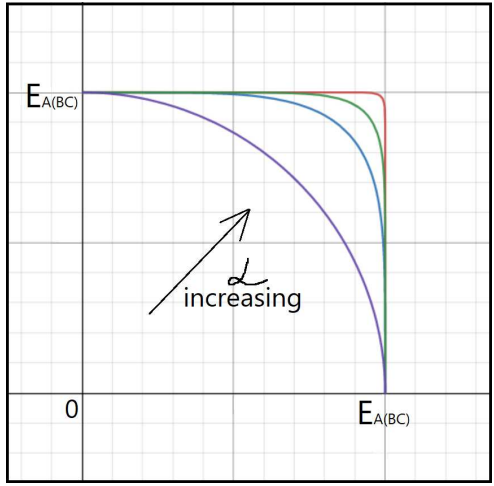}
\caption{Curves for $\alpha=2, 10, 15, 50$.}
\end{figure} 
\end{center}

So now in order to make the Definition 16 equivalent to the Definition 15, we need to make only one change, which will be highlighted with a bold text below.

\begin{definition}
If there exists a function $f:R\times R\rightarrow R$ (where $R$ is the real numbers in $[0,\infty)$ interval) such that 

\begin{equation}
E_{A(BC)}(\rho_{ABC})\geq f(E_{AB}(\rho_{AB}),E_{AC}({\rho_{AC}}))\geq\max(E_{AB}(\rho_{AB}),E_{AC}({\rho_{AC}}))
\end{equation}

is true \textbf{for every fixed dimension $\mathbf{d=dim(\mathcal{H}_A\otimes\mathcal{H}_B\otimes\mathcal{H}_C)}$} and all tripartite states $\rho_{ABC}$ in a composite system with Hilbert space $\mathcal{H}_A\otimes\mathcal{H}_B\otimes\mathcal{H}_C$, and such that 

\begin{equation}
f(E_{AB}(\rho_{AB}),E_{AC}({\rho_{AC}}))=\max(E_{AB}(\rho_{AB}),E_{AC}(\rho_{AC}))
\end{equation}

is only true at $(E_{AB}(\rho_{AB}),E_{AC}(\rho_{AC}))=(0,E_{A(BC)}(\rho_{ABC}))$ and $(E_{AB}(\rho_{AB}),E_{AC}(\rho_{AC}))=$ \newline$=(E_{A(BC)}(\rho_{ABC}),0)$, then the entanglement measure $E$ is monogamous. 
\end{definition}

So, we again alter the definition so that we don't consider infinite dimensions of the tripartite states $\rho_{ABC}$. Remember a similar consideration was made in the subchapter 4.1 above the inequality (45), when the Definition 14 failed monogamy for the entanglement of formation and relative entropy of entanglement. Now let us see that this definition is indeed the same as the Definition 15. This then will lead us to the proof that the inequalities (45) and (46) show that the entanglement of formation and the regularised relative entropy are monogamous in the sense of the Definition 15.

\textbf{Proof that (45) and (46) display monogamy of the Definition 15.} The equivalence of the Definition 17 and the Definition 15 is true because of the Theorem 4. It is easy to see that the definition of monogamy given inside the Theorem 4 is identical to the Definition 17, except that in the Definition 17 we have some function $f$, while in the Theorem 4 we have $(E_{AB}(\rho_{AB})^{\alpha}+E_{AC}(\rho_{AC})^{\alpha})^{1/\alpha}$ for some $\alpha$. Notice that $f$ and $\alpha$ are chosen for a fixed dimension of tripartite states in both cases. While reading the proof it is advised to compare the Figures 5 and 9 for visualisation. So, to show equivalence between the Definition 17 and the Definition 15 we need the following two-way proof:

\begin{adjustwidth*}{0em}{2em}

$(\implies)$ Suppose an entanglement measure $E$ is monogamous in the sense of the Definition 15. Then let us simply choose $f=(E_{AB}(\rho_{AB})^{\alpha}+E_{AC}(\rho_{AC})^{\alpha})^{1/\alpha}$ (by the Theorem 4). Note that we have $(E_{AB}(\rho_{AB})^{\alpha}+E_{AC}(\rho_{AC})^{\alpha})^{1/\alpha}\geq\max(E_{AB}(\rho_{AB}),E_{AC}(\rho_{AC}))$  for all $E_{AB}$, $E_{AC}\in [0,E_{A(BC)}]$ and all $\alpha>0$. Thus, the entanglement measure $E$ is monogamous in the sense of the Definition 17.

$(\impliedby)$ Suppose an entanglement measure $E$ is monogamous in the sense of the Definition 17. Then for every curve 

\begin{equation}
E_{A(BC)}(\rho_{ABC})=f(E_{AB}(\rho_{AB}),E_{AC}({\rho_{AC}}))
\end{equation}

which is entirely inside the interior of the square defined by (48), except for its two endpoints at $(E_{A(BC)}(\rho_{ABC},0)$ and $(0,E_{A(BC)}(\rho_{ABC})$, we can find an $\alpha>0$ such that 

\begin{equation}
f(E_{AB}(\rho_{AB}),E_{AC}({\rho_{AC}}))\geq (E_{AB}(\rho_{AB})^{\alpha}+E_{AC}(\rho_{AC})^{\alpha})^{1/\alpha}
\end{equation}

is true for all $E_{AB}$, $E_{AC}\in [0,E_{A(BC)}]$, therefore, the entanglement measure $E$ is monogamous in terms of the Definition 15 (by the Theorem 4).

The inequality (55) can be satisfied for some $\alpha>0$ because for every point $(E_{AB},E_{AC})$ inside the interior of the square defined by (48) we can find $\alpha>0$ such that the shape defined by inequality (51) engulfs this point. This is true because $(x^{\alpha}+y^{\alpha})^{1/\alpha}\rightarrow \max(x,y)$ as $\alpha\rightarrow\infty$ \textit{[See \textbf{Proof 14} in the Appendix]}. And the curve (54) is entirely made up of these interior points, except for its endpoints. Thus, we can find $\alpha>0$ such that the shape defined by (51) engulfs that curve, meaning that the inequality (55) is satisfied. 

\end{adjustwidth*}

\hfill $\square$

This is the original observation of this dissertation. We showed that the inequalities (45) and (46) from the previous subchapter (which were deduced in [28]) prove that the entanglement of formation and the regularised relative entropy of entanglement are monogamous in the sense of the Definition 15. This is because they are clearly monogamous in the sense of the Definition 17. And we have just proved that these two definitions are equivalent. 

Authors of the paper [41], who introduced the Definition 15, seem to have made a slight oversight on this. This is because they only mentioned that the entanglement of formation has been proven to be non-monogamous in the sense of the Definition 14 (where infinite dimensions are considered). But they never talked about the inequalities (45) and (46) possibly displaying monogamous properties of their respective entanglement measures. They instead came up with an entirely different proof in [42] of the fact that the entanglement of formation is monogamous in the sense of the Definition 15.

Thus far, we have observed by how much both of the definitions of monogamous entanglement measures from [28] and [41] differ from each other. We have done it by gradually making the definition from [28] becoming equivalent to the definition from [41]. And we indeed have finally achieved it after the introduction of the Theorem 4. 

But we still do not know whether the definition of monogamous entanglement measures was better in [28] or in [41]. It does indeed seem more natural that entanglement should have a property such that if 

\begin{equation}
E_{A(BC)}(\rho_{ABC})=E_{AB}(\rho_{AB})
\end{equation}

then we have $E_{AC}(\rho_{AC})=0$. But there is still no solid mathematical proof for this. Indeed, what if there does exist a tripartite state such that there is still some entanglement between the particles in systems A and C in this case? This question remains unresolved. 

\large
\subsection{Further motivation for studying monogamy}

\normalsize

In this dissertation we were only focusing on studying monogamy just as another quantum entanglement property that must be satisfied by "good" entanglement measures. But this is obviously not the only reason to investigate it. There exist a lot of problems concerning real physical phenomena where monogamy was being considered when working on them. For instance, these include: the security of quantum key distribution [46,47], condensed-matter physics [48], frustrated spin systems [49] and black-hole physics [50]. For example, in [50] it was argued that the black hole evaporation (discovered by Stephen Hawking [51]) is incompatible with our understanding of monogamy of entanglement. These claims were further examined in papers like [52,53]. 

There already exist numerous papers where monogamy is contemplated in order to solve some of the existing problems in physics. It is only a matter of time when new problems appear that will too require a sufficient knowledge about monogamy of entanglement. Thus, it is very important to sharpen our understanding of monogamy further. 

\begin{center}
\Large
\section{Conclusion}
\end{center}

\normalsize
In this dissertation we have had a brief and informative introduction to the entanglement quantification and an in-depth examination of the entanglement property, called monogamy. 

We explained why each of the properties that we presented must be satisfied by a "good" entanglement measures. These properties included: no entanglement for separable states, entanglement does not increase under LOCC operations on average, a maximally entangled state is given by (25), additivity, convexity, continuity. 

We have confirmed that the Von Neumann entropy of entanglement is rightfully a universally accepted entanglement measure for pure states by showing that it follows all of the entanglement properties. 

We introduced some of the most prevalent entanglement measures that have been proposed so far: entanglement of formation, relative entropy of entanglement, squashed entanglement, entanglement cost, distillation of entanglement. We have listed a plethora of reasons of why none of these entanglement measures are flawless. Either the entanglement measure does not follow one of the entanglement properties, or its value still remains too difficult to compute for an arbitrary mixed state. 

Finally, we were introduced to the seventh entanglement property. The property which does not allow limitless distribution of entanglement across many subsystems. Specifically, we focused on how entanglement is shared across three subsystems. 

We looked at one of the first mathematical representations that displayed monogamous properties for an entanglement measure called the squared concurrence. This mathematical representation involved inequalities which were comparing amount of entanglement between three subsystems. We then noted that specifically this kind of inequality is not satisfied by entanglement measures such as entanglement of formation and the distillable entanglement. Though next we wondered if these entanglement measures could still somehow exhibit monogamous properties through a different inequality, but with a similar concept behind it. 

We looked at an attempt which tried to see if the entanglement of formation and the relative entropy of entanglement are monogamous in terms of the Definition 14. We showed that they are not. But then after revising the Definition 14 to be less demanding, we showed that the inequalities (45) and (46) are true for the entanglement of formation and the relative entropy of entanglement. And we explained that they have just as much right to be displaying monogamy of an entanglement measure as the inequality (34) has. 

Then we looked at another attempt of defining monogamous entanglement measures. This was done with the Definition 15, which used equalities instead. We made comparisons of this definition with the previous one. We mentioned that there was an entire paper dedicated to the proof that the entanglement of formation is monogamous in terms of the Definition 15. However, I was able to make an original observation and show that the inequality (45), which existed before the publication of that paper, already served as a proof of the fact that the entanglement of formation is monogamous in the sense of the Definition 15. The same observation also automatically led us to the revelation that the inequality (46) proves that the regularised relative entropy of entanglement is monogamous in the sense of the Definition 15 as well. Unlike the proof of the monogamy of the entanglement of formation, this fact has not been proven for the regularised relative entropy of entanglement before. 

There has been done an enormous amount of work by lots of experts in the field. Numerous entanglement measures have been proposed and many hours have been put into the research of monogamy of entanglement. However, these and other areas of quantum mechanics still require a lot more work to be done in order for us to have a sufficient understanding of the physical phenomenon that has been perplexing scientists for nearly one hundred years.

\bigskip

\begin{figure}[h]
\centering
\section{Appendix}
\end{figure}

\normalsize

In this section I present some of the proofs of the statements that were made during this dissertation. Specifically, I will be aiming to mostly give much more detailed proofs for the claims that did not have a rigorous enough proof in their original papers. 

\begin{proof}
To see why this is the case, one would only need to check that "the spooky action at a distance" does not arise if we were to perform a similar calculation as in the EPR paradox but with this state instead. So suppose we perform a quantum measurement on a particle in system $A_1$ with the possible outcomes $\{\lambda_1, ... , \lambda_k\}$ and PVM $\{P_1\otimes I^{\otimes (n-1)}, ... , P_k\otimes I^{\otimes (n-1)}\}$, then the residual joint state if we get the outcome $\lambda_m$ will be

\begin{equation}
\rho_{A_1...A_n}=\sum_i{p_i\left(\frac{P_m\rho^{A_1}_i P_m}{Tr(P_m\rho^{A_1}_i P_m)}\otimes...\otimes\rho^{A_n}_i\right)}
\end{equation}

then the joint state of the particles in composite system with Hilbert space $\mathcal{H}_{A_2}\otimes...\otimes\mathcal{H}_{A_n}$ is

\begin{equation} 
\begin{split}
&Tr_{A_1}\left(\sum_i{p_i\left(\frac{P_m\rho^{A_1}_i P_m}{Tr(P_m\rho^{A_1}_i P_m)}\otimes...\otimes\rho^{A_n}_i\right)}\right)=\\
&=\sum_i{p_i Tr_{A_1}\left(\frac{P_m\rho^{A_1}_i P_m}{Tr(P_m\rho^{A_1}_i P_m)}\otimes...\otimes\rho^{A_n}_i\right)}=\\
&=\sum_i{p_i \left(\frac{Tr(P_m\rho^{A_1}_i P_m)}{Tr(P_m\rho^{A_1}_i P_m)} \rho^{A_2}_i \otimes...\otimes\rho^{A_n}_i\right)}=\\
&=\sum_i{p_i (\rho^{A_2}_i \otimes...\otimes\rho^{A_n}_i)}
\end{split}
\end{equation} 

while the joint state of the same system before the measurement was

\begin{equation} 
\begin{split}
Tr_{A_1}\left(\sum_i{p_i (\rho^{A_1}_i \otimes...\otimes\rho^{A_n}_i)}\right)&=\sum_i{p_i (Tr(\rho^{A_1}_i)\rho^{A_2}_i \otimes...\otimes\rho^{A_n}_i)}=\\
&=\sum_i{p_i (\rho^{A_2}_i \otimes...\otimes\rho^{A_n}_i)}
\end{split}
\end{equation} 

where $Tr(\rho^{A_1}_i)=1$ because $\rho^{A_1}_i$ is a density matrix. Thus, indeed the joint state of the particles in the composite system with Hilbert space $\mathcal{H}_{A_2}\otimes...\otimes\mathcal{H}_{A_n}$ is unchanged regardless of the outcome of the measurement on the particle in the system with Hilbert space $\mathcal{H}_{A_1}$. 

For a pure separable state, one would have to do the identical operations to confirm that it has no entanglement. 

\hfill $\square$

\end{proof} 

\begin{proof}
To see that this is the maximally entangled state we need to show that we can transform it into any state we want. This will be a valid proof because of the second property. Therefore, the only state that can get transformed with LOCC into any other state must be maximally entangled. 

\begin{center}
\begin{figure}[h]
\centering
\includegraphics[scale=0.42]{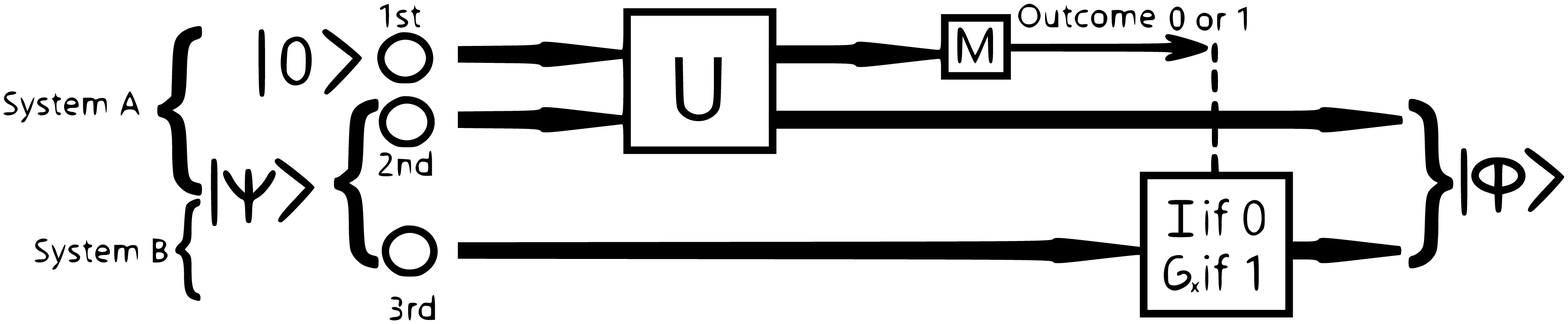}
\caption{LOCC for $\ket{\psi}\rightarrow\ket{\phi}$.}
\end{figure} 
\end{center}

Now to prove that this is true let us just consider the case for $d=2$ for simplicity (the generalisation for arbitrary d is done similarly). First let us prove that we can transform our state into $\ket{\phi}=\alpha\ket{00}+\beta\ket{11}$. Suppose we have three particles. The first one is in system A and is in state $\ket{0}$, the second and the third are in system A and B respectively and are in the joint state $\ket{\psi}=\frac{\ket{0}\otimes\ket{0}+\ket{1}\otimes\ket{1}}{\sqrt{2}}$. Then we would apply a unitary U on the first and second particle such that $\ket{00}\rightarrow\alpha\ket{00}+\beta\ket{11}$ and $\ket{01}\rightarrow\alpha\ket{01}+\beta\ket{10}$. So the joint state of the three particles is 

\begin{equation}
\frac{\ket{0}(\alpha\ket{00}+\beta\ket{11})+\ket{1}(\alpha\ket{10}+\beta\ket{01})}{\sqrt{2}}
\end{equation}

then there would be a measurement performed on the first particle with PVM $\{\ket{0}\bra{0}\otimes I\otimes I, \ket{1}\bra{1}\otimes I\otimes I\}$. If outcome is $0$ then do nothing, if the outcome is $1$ then apply $\sigma_x$ Pauli matrix on the third particle. We just showed that with LOCC we can transform $\ket{\psi}$ any pure state because any pure state can be written as a Schmidt decomposition like $\ket{\phi}$. 

Then it also follows that we can transform $\ket{\psi}$ into any mixed state because any mixed state can be written as $\rho=\sum_i{p_i\ket{\phi_i}\bra{\phi_i}}$ where $\ket{\phi_i}=U_i\otimes V_i(\alpha_i\ket{00}+\beta_i\ket{11})$ with unitaries $U_i$ and $V_i$. To make $\ket{\psi}\rightarrow\ket{\phi_i}$ we have to do the same operations as before but then also apply $U_i\otimes V_i$ to the second and third particle. To get $\ket{\psi}\rightarrow\rho$ we need to make sure that the state $\ket{\phi_i}$ is prepared with probability $p_i$ in our experiment, which will create the statistical mixture $\rho$. 

\hfill $\square$

\end{proof}

\begin{proof}

We will show that each property from subsection 3.1 is satisfied for the Von Neumann entropy of entanglement:

\textbf{Property 1}

We will show that the Von Neumann entropy of a separable state indeed gives us the lowest value possible. If $\ket{\psi}=\ket{\psi_A}\otimes\ket{\psi_B}$ then

\begin{equation}
S(\ket{\psi})=-Tr(\ket{\psi_A}\bra{\psi_A}\log_2(\ket{\psi_A}\bra{\psi_A}))=-Tr(\ket{\psi_A}\bra{\psi_A}\ket{\psi_A}\bra{\psi_A}\log_2(1))=0
\end{equation}

where we used $f(N)=\sum_{\lambda\in\sigma(N)}f(\lambda) P_{\lambda}$ for normal operators $N$. 

\textbf{Property 2}

It was proven in [7] that the Von Neumann entropy is non-increasing under LOCC on average. 

\textbf{Property 3}

We will show that the Von Neumann entropy of a maximally entangled sate has the maximum value. So let us first compute the Von Neumann entropy of the maximally entangled state $\ket{\psi}=\frac{\ket{0}\otimes\ket{0}+...+\ket{d-1}\otimes\ket{d-1}}{\sqrt{d}}$ of two particles, each in d dimensional sate:

\begin{equation}
\begin{split}
\rho_A=Tr_B(\ket{\psi}\bra{\psi})&=Tr_B\left(\frac{1}{d}\sum_{i=0,j=0}^{d-1}{(\ket{i}\otimes\ket{i})(\bra{j}\otimes\bra{j})}\right)=\\
&=Tr_B\left(\frac{1}{d}\sum_{i=0,j=0}^{d-1}{\ket{i}\bra{j}\otimes\ket{i}\bra{j}}\right)=\\
&=\frac{1}{d}\sum_{i=0,j=0}^{d-1}{\ket{i}\bra{j}Tr(\ket{i}\bra{j}})=\frac{1}{d}\sum_{i=0,j=0}^{d-1}{\ket{i}\bra{j}\delta_{ij}}=\frac{1}{d}\sum_{i=0}^{d-1}{\ket{i}\bra{i}}
\end{split}
\end{equation}

therefore we have 

\begin{equation}
\begin{split}
S(\ket{\psi})&=-Tr(\rho_A \log_2(\rho_A))=\\
&=-Tr\left(\frac{1}{d}\sum_{i=0}^{d-1}{\ket{i}\bra{i}} \log_2\left(\frac{1}{d}\sum_{i=0}^{d-1}{\ket{j}\bra{j}}\right)\right)=\\
&=-Tr\left(\frac{1}{d}\sum_{i=0,j=0}^{d-1}{\ket{i}\bra{i}\ket{j}\bra{j}} \log_2\left(\frac{1}{d}\right)\right)=\\
&=-Tr\left(\frac{1}{d}\sum_{i=0}^{d-1}{\ket{i}\bra{i}} \log_2\left(\frac{1}{d}\right)\right)=\\
&=-Tr(I)\frac{1}{d} \log_2\left(\frac{1}{d}\right)=-\log_2\left(\frac{1}{d}\right)\\
\end{split}
\end{equation}

where $I$ is a d-dimensional identity matrix, therefore $-Tr(I)=d$.

Now suppose we have a state with Schmidt decomposition  $\ket{\phi}=\sum_{i=0}^{d-1}{\sqrt{\mu_i}\ket{e_i}\otimes\ket{f_i}}$ where $\mu_i\geq 0$ and $\sum_{i=0}^{d-1}{\mu_i}=1$. Notice we set the Schmidt coefficient to $r=d-1$ because we allow some $\mu_i=0$. So let us compute the Von Neumann entropy of the general pure state $\ket{\phi}$:

\begin{equation}
\begin{split}
\rho_A=Tr_B(\ket{\phi}\bra{\phi})&=Tr_B\left(\sum_{i=0,j=0}^{d-1}{\sqrt{\mu_i}\sqrt{\mu_j}(\ket{e_i}\otimes\ket{f_i})(\bra{e_j}\otimes\bra{f_j})}\right)=\\
&=\sum_{i=0,j=0}^{d-1}{\sqrt{\mu_i}\sqrt{\mu_j}\ket{e_i}\bra{e_j} Tr(\ket{f_i}\bra{f_j})}=\\
&=\sum_{i=0,j=0}^{d-1}{\sqrt{\mu_i}\sqrt{\mu_j}\ket{e_i}\bra{e_j}\delta_{ij}}=\sum_{i=0}^{d-1}{\mu_i\ket{e_i}\bra{e_i}}
\end{split}
\end{equation}

therefore we have 

\begin{equation}
\begin{split}
S(\ket{\phi})&=-Tr\rho_A \log_2(\rho_A))=\\
&=-Tr\left(\sum_{i=0}^{d-1}{\mu_i\ket{e_i}\bra{e_i}} \log_2\left(\sum_{i=0}^{d-1}{\mu_j\ket{e_j}\bra{e_j}}\right)\right)=\\
&=-Tr\left(\sum_{i=0,j=0}^{d-1}{\mu_i\ket{e_i}\bra{e_i}\ket{e_j}\bra{e_j}} \log_2(\mu_j)\right)=\\
&=-Tr\left(\sum_{i=0}^{d-1}{\mu_i\ket{e_i}\bra{e_i}} \log_2(\mu_i)\right)=-\sum_{i=0}^{d-1}{\mu_i \log_2(\mu_i)}
\end{split}
\end{equation}

Then we can see that

\begin{equation}
-\log_2\left(\frac{1}{d}\right)=-\sum_{i=0}^{d-1}{\frac{1}{d}\log_2\left(\frac{1}{d}\right)}\geq-\sum_{i=0}^{d-1}{\mu_i \log_2(\mu_i)}
\end{equation}

where $\sum_{i=0}^{d-1}{\frac{1}{d}}=\sum_{i=0}^{d-1}\mu_i=1$. To see why the last inequality is true one has to differentiate 

\begin{equation}
-\sum_{i=0}^{d-1}{\mu_i \log_2(\mu_i)}=-\sum_{i=0}^{d-2}{\mu_i \log_2(\mu_i)}-(1-\sum_{m=0}^{d-2}{\mu_m})\log_2(1-\sum_{n=0}^{d-2}{\mu_n})
\end{equation}

with respect to each $\mu_i$ for $i=0,...,d-2$, equate to zero, then see that this multivariate function achieves maximum value when $\mu_i=\frac{1}{d}$.

We have used $f(N)=\sum_{\lambda\in\sigma(N)}f(\lambda) P_{\lambda}$ for normal operators $N$ two times in this proof. 

\textbf{Property 4}

To see that the Von Neumann entropy of entanglement is additive, $S(\ket{\psi}^{\otimes n})=nS(\ket{\psi})$, one needs to note that 

\begin{equation}
\begin{split}
&Tr_B((\ket{\psi}\bra{\psi})^{\otimes n})=Tr_B\left(\left(\sum_{i,j}{\sqrt{\mu_i}\sqrt{\mu_j}\ket{e_i}\bra{e_j}\otimes\ket{f_i}\bra{f_j}}\right)^{\otimes n}\right)=\\
&=Tr_B\left(\sum_{i_1...i_n}\sum_{j_1...j_n}\sqrt{\mu_{i_1}...\mu_{i_n}}\sqrt{\mu_{j_1}...\mu_{j_n}}\ket{e_{i_1}...e_{i_n}}\bra{e_{j_1}...e_{j_n}}\otimes\ket{f_{i_1}...f_{i_n}}\bra{f_{j_1}...f_{j_n}}\right)=\\
&=\sum_{i_1...i_n}\sum_{j_1...j_n}\sqrt{\mu_{i_1}...\mu_{i_n}}\sqrt{\mu_{j_1}...\mu_{j_n}}\ket{e_{i_1}...e_{i_n}}\bra{e_{j_1}...e_{j_n}}Tr(\ket{f_{i_1}...f_{i_n}}\bra{f_{j_1}...f_{j_n}})=\\
&=\sum_{i_1...i_n}\sum_{j_1...j_n}\sqrt{\mu_{i_1}...\mu_{i_n}}\sqrt{\mu_{j_1}...\mu_{j_n}}\ket{e_{i_1}...e_{i_n}}\bra{e_{j_1}...e_{j_n}}\delta_{i_1j_1}...\delta_{i_nj_n}=\\
&=\sum_{i_1...i_n}\mu_{i_1}...\mu_{i_n}\ket{e_{i_1}...e_{i_n}}\bra{e_{i_1}...e_{i_n}}
\end{split}
\end{equation}

then one would need to make an observation that

\begin{equation}
\begin{split}
\log_2(Tr_B(\ket{\psi}\bra{\psi})^{\otimes n})&=\log_2\left(\sum_{i_1, ... , i_n}{\mu_{i_1}...\mu_{i_n}\ket{e_{i_1}...e_{i_n}}\bra{e_{i_1}...e_{i_n}}}\right)=\\
&=\sum_{i_1, ... , i_n}{\log_2(\mu_{i_1}...\mu_{i_n})\ket{e_{i_1}...e_{i_n}}\bra{e_{i_1}...e_{i_n}}}
\end{split}
\end{equation}

where we once again used $f(N)=\sum_{\lambda\in\sigma(N)}f(\lambda) P_{\lambda}$ for normal operators $N$. Now all that's left is to add these values into $S(\ket{\psi}^{\otimes n})=-Tr(Tr_B(\ket{\psi}\bra{\psi})^{\otimes n}) \log_2(Tr_B(\ket{\psi}\bra{\psi})^{\otimes n}))$ to get 

\begin{equation}
\begin{split}
-&Tr\left(\sum_{i_1, ... , i_n}{\mu_{i_1}...\mu_{i_n}\ket{e_{i_1}...e_{i_n}}\bra{e_{i_1}...e_{i_n}}}\sum_{j_1, ... , j_n}{\log_2(\mu_{j_1}...\mu_{j_n})\ket{e_{j_1}...e_{j_n}}\bra{e_{j_1}...e_{j_n}}}\right)=\\
&=-Tr\left(\sum_{i_1, ... , i_n}{\mu_{i_1}...\mu_{i_n}\ket{e_{i_1}...e_{i_n}}\bra{e_{j_1}...e_{j_n}}\delta_{i_1 j_1}...\delta_{i_n j_n}}\sum_{j_1, ... , j_n}{\log_2(\mu_{j_1}...\mu_{j_n})}\right)=\\
&=-Tr\left(\sum_{i_1, ... , i_n}{\mu_{i_1}...\mu_{i_n}\ket{e_{i_1}...e_{i_n}}\bra{e_{i_1}...e_{i_n}}}{\log_2(\mu_{j_1}...\mu_{j_n})}\right)=\\
&=\sum_{i_1, ... , i_n}{\mu_{i_1}...\mu_{i_n}}{\log_2(\mu_{i_1}...\mu_{i_n})}=\sum_{i_1, ... , i_n}{\mu_{i_1}...\mu_{i_n}}(\log_2(\mu_{i_1})+...+\log_2(\mu_{i_n}))=\\
&=\sum_{i_1, ... , i_n}{\mu_{i_1}...\mu_{i_n}}\log_2(\mu_{i_1})+...+\sum_{i_1, ... , i_n}{\mu_{i_1}...\mu_{i_n}}\log_2(\mu_{i_n})=\\
&=n\sum_{i_1, ... , i_n}{\mu_{i_1}...\mu_{i_n}}\log_2(\mu_{i_1})=n\sum_{i_1}{\mu_{i_1}\log_2(\mu_{i_1})\sum_{i_2, ... , i_n}\mu_{i_2}...\mu_{i_n}}=\\
&=n\sum_{i_1}{\mu_{i_1}\log_2(\mu_{i_1})\prod_{i_m=i_2}^{i_n}\sum_{i_m}\mu_{i_m}}=n\sum_{i_1}\mu_{i_1}\log_2(\mu_{i_1})\prod_{i_m=i_2}^{i_n}{1}\\
&=n\sum_{i_1}\mu_{i_1}\log_2(\mu_{i_1})=n S(\ket{\psi})
\end{split}
\end{equation}

\textbf{Property 5}

$S(\sum_k{p_k\rho_k})\geq \sum_k{p_k S(\rho_k)}$ for $\rho=\sum_k p_k\rho_k$ has been shown to be true in [16]. 

\textbf{Property 6}

The continuity was proven in [8] by deriving the Fannes' inequality

\begin{equation}
\abs{S(\rho)-S(\sigma)}\leq T(\rho,\sigma) \log(d)+\eta(T(\rho,\sigma))
\end{equation} 

for all states $\rho,\sigma\in \mathcal{H}$, where $d$ is the dimension of the Hilbert $\mathcal{H}$, $\eta(x)=-x\log(x)$ and $T(\rho,\sigma)=\frac{1}{2}Tr(\sqrt{(\rho-\sigma)^{\dagger}(\rho-\sigma)})$. One simply needs to see that the right hand side of the inequality (71) equals to zero when $\rho=\sigma$ to see that continuity is satisfied. 

\hfill $\square$

\end{proof}

\begin{proof}

We will show that each property (except for additivity) from subsection 3.1 is satisfied for the entanglement of formation: 

\textbf{Property 1}

We will show that the entanglement of formation of a separable state $\rho_{AB}$ is zero. 

\begin{equation}
\begin{split}
E_F(\sum_i{p_i \rho^A_i\otimes \rho^B_i})&=E_F(\sum_{ijk}{p_i q_j r_k \ket{\psi_{ij}}\bra{\psi_{ij}}\otimes \ket{\phi_{ik}}\bra{\phi_{ik}}})=\\
&=E_F(\sum_{ijk}{p_i q_j r_k \ket{\psi_{ij}\otimes\phi_{ik}}\bra{\psi_{ij}\otimes\phi_{ik}}})=\\
&=\min_{\{\alpha_n,\ket{\omega_n}\}}\sum_n\alpha_n S(\ket{\omega_n}\bra{\omega_n})
\end{split}
\end{equation}

such that $\rho_{AB}=\sum_{n}\alpha_n\ket{\omega_n}\bra{\omega_n}$ and where we used the fact that every density matrix $\rho$ can be written as $\sum_i p_i \ket{\psi_i}\bra{\psi_i}$ such that $p_i>0$ (which follows from the spectral theorem and the properties of a density matrix). Although it turns out that the decomposition that we have done for the state $\rho_{AB}$ is a minimal one already:

\begin{equation}
\sum_{ijk}{p_i q_j r_k S(\ket{\psi_{ij}\otimes\phi_{ik}}\bra{\psi_{ij}\otimes\phi_{ik}})}=0
\end{equation}

Here we used the result from the Proof 3 where for the Property 1 we showed that the Von Neumann entropy of a pure separable state is zero. And zero is the minimal value of the entanglement of formation because the Von Neumann entropy is greater or equal to zero for all pure states. 

\textbf{Property 2}
Non-increasing expectation of entanglement of formation under LOCC has been proven in [17] (in sub-chapter 2.1).

\textbf{Property 3}

It is straightforward that entanglement of formation $E_F(\rho)$ is equal to the Von Neumann entropy if the state $\rho$ is pure. Therefore, since the maximally entangled state $\ket{\psi}\bra{\psi}$ of two particles, each in d-dimensional state, is pure, we get 

\begin{equation}
E_F(\ket{\psi}\bra{\psi})=S(\ket{\psi}\bra{\psi})=-\log_2\left(\frac{1}{d}\right)\geq-\sum_{i=0}^{d-1}{\mu_i \log_2(\mu_i)}
\end{equation}

where we have used the results from the Proof 3 of the third property, where we were doing Schmidt decompositions and $\mu_i$ were the Schmidt coefficients. While the entanglement of formation of an arbitrary state $\rho$ can be written as 

\begin{equation}
\begin{split}
&E_F(\rho)=\sum_i{p_iS(\ket{\psi_i}\bra{\psi_i})}=\sum_i{p_i\left(-\sum_j{\mu_j \log_2(\mu_j)}\right)_i}\leq\\
&\leq \sum_i{p_i\left(-\log_2\left(\frac{1}{d}\right)\right)_i}=-\log_2\left(\frac{1}{d}\right)\sum_i{p_i}=-\log_2\left(\frac{1}{d}\right)=E_F(\ket{\psi}\bra{\psi})
\end{split}
\end{equation}

\textbf{Property 4}

As mentioned before, it was proven in [15] entanglement of formation doesn't follow the property of additivity.

\textbf{Property 5}

Convexity automatically follows since the Von Neumann entropy of entanglement is convex. And the sum of convex functions is a convex function.

\textbf{Property 6}

The continuity of the entanglement of formation has been proven in [43] where it was deduced that 

\begin{equation}
E_F(\rho)-E_F(\sigma)\leq (5\log(d)+4\log(d'))D(\rho,\sigma)+2\eta (D(\rho,\sigma))
\end{equation}

for all $\rho$ and $\sigma$, where $D(\rho,\sigma)=2\sqrt{1-Tr(\sqrt{\rho^{1/2}\sigma\rho^{1/2}})}$ and $d,d'$ are the respective dimensions of these states. Here once again has to only note that the right hand side is equal to zero when $\rho=\sigma$ and that trace of a density matrix is zero to see that continuity is satisfied. 

\hfill $\square$

\end{proof}

\begin{proof}

Here we will only give a summary of the large and complex development of the proof that spans over the papers [17-19]. 

In the first paper [17] the "large proof" begins with the derivation of the Von Neumann entropy of entanglement of a pure state $\ket{\psi}$ describing two qubits, when we know its decomposition in terms of the Bell ONB. So if $\ket{\phi}=\sum_{i=1}^4{\alpha_i\ket{e_i}\bra{e_i}}$ where $\ket{e_i}$ stands for a Bell state then then one can directly compute that 

\begin{equation}
S(\ket{\psi})=-\frac{1+\sqrt{1-C^2}}{2}\log_2\left(\frac{1+\sqrt{1-C^2}}{2}\right)-\frac{1-\sqrt{1-C^2}}{2}\log_2\left(\frac{1-\sqrt{1-C^2}}{2}\right)
\end{equation}

where $C=\abs{\sum_i{\alpha_i^2}}$. From the very beginning we have a function that looks exactly like the result we are going towards, except $C$ appears to be different. However, it turns out that the concurrence $C=\max\{0, \lambda_1-\lambda_2-\lambda_3-\lambda_4\}$ that we defined earlier reduces to $C=\abs{\sum_i{\alpha_i^2}}$ for pure states. 

Then in the second paper [18] it was proven that the entanglement of formation of a mixed state $\rho$ describing two qubits is given by 

\begin{equation}
E_F(\rho)=-\frac{1+\sqrt{1-C^2}}{2}\log_2\left(\frac{1+\sqrt{1-C^2}}{2}\right)-\frac{1-\sqrt{1-C^2}}{2}\log_2\left(\frac{1-\sqrt{1-C^2}}{2}\right)
\end{equation}

if $\rho$ has at most two non-zero eigenvalues. Here $C$ again stands for the concurrence. Also, this paper provider evidence that the formula could work for any mixed state $\rho$. This gave motivation for the development of the proof of the final result, which indeed has been achieved in [19]. 

\hfill $\square$

\end{proof}

\begin{proof}
If we have a pure entangled state $\ket{\psi}$ of composite system with Hilbert space $\mathcal{H}_A\otimes\mathcal{H}_B$ then it can't be written as a product state and therefore its only possible Schmidt decompositions are

\begin{equation}
\ket{\psi}=\sum_i^r{\sqrt{\mu_i}\ket{e_i}\otimes\ket{f_i}}
\end{equation}

such that $r\geq2$. Then if we take a partial trace of the joint state then we get 

\begin{equation}
Tr_B(\ket{\psi}\bra{\psi})=Tr_B(\sum_{ij}^r{\sqrt{\mu_i}\sqrt{\mu_j}\ket{e_i}\bra{e_j}\otimes\ket{f_i}\bra{f_j}})=\sum_{i}^r\mu_i\ket{e_i}\bra{e_j}
\end{equation}

which is indeed a mixed state since $r\geq 2$.

\hfill $\square$

\end{proof}

\begin{proof}

To prove that the inequality (34) is true for any tripartite pure state we begin by reminding ourselves of the definition of concurrence. Concurrence of a bipartite state $\rho$ of two qubits is given by $C=\max\{0, \lambda_1-\lambda_2-\lambda_3-\lambda_4\}$, where $\lambda_i$ are the eigenvalues of $R=\sqrt{\rho\tilde{\rho}}$ in descending order, and where $\tilde{\rho}=(\sigma_y\otimes\sigma_y)\rho^*(\sigma_y\otimes\sigma_y)$ with $\sigma_y$ being the y-Pauli matrix. 

Then we note that if $\rho$ is a pure state then $C=2\sqrt{Det(\rho_A)}=2\sqrt{Det(\rho_B)}$, where $\rho_A$ and $\rho_B$ are the partial states of each of the two qubits. 

\begin{adjustwidth*}{2em}{0em}

\textbf{Subproof 1} To prove this, we must find the eigenvalues of $R$, so let us start by writing our bipartite state as

\begin{equation}
\begin{split}
\rho&=\ket{\psi}\bra{\psi}=\\
&=(a\ket{00}+b\ket{01}+c\ket{10}+d\ket{11})(a^*\bra{00}+b^*\bra{01}+c^*\bra{10}+d^*\bra{11})
\end{split}
\end{equation}

where we have a general state of $\ket{\psi}=a\ket{00}+b\ket{01}+c\ket{10}+d\ket{11}$. Though this time it will actually be easier to work with matrices instead of the bra/ket notations: 

\begin{equation}
\rho=\begin{pmatrix}
           a \\
           b \\
           c \\
           d
         \end{pmatrix}(a^*,b^*,c^*,d^*)=\begin{pmatrix}
aa^*&ab^*&ac^*&ad^* \\
ba^*&bb^*&bc^*&bd^* \\
ca^*&cb^*&cc^*&cd^* \\
da^*&db^*&dc^*&dd^*
\end{pmatrix}
\end{equation}

then lets us transform the tensor product of y-Pauli matrices as well:

\begin{equation}
\sigma_y\otimes\sigma_y=\begin{pmatrix}
0&-i \\
i&0 \\
\end{pmatrix}\otimes\begin{pmatrix}
0&-i \\
i&0 \\
\end{pmatrix}=\begin{pmatrix}
0&0&0&-1 \\
0&0&1&0 \\
0&1&0&0 \\
-1&0&0&0
\end{pmatrix}
\end{equation}

therefore we get the value of $\tilde{\rho}$ to be

\begin{equation}
\begin{split}
\tilde{\rho}&=(\sigma_y\otimes\sigma_y)\rho^*(\sigma_y\otimes\sigma_y)=(\sigma_y\otimes\sigma_y)\begin{pmatrix}
-ad^*&ac^*&ab^*&-aa^* \\
-bd^*&bc^*&bb^*&-ba^* \\
-cd^*&cc^*&cb^*&-ca^* \\
-dd^*&dc^*&db^*&-da^*
\end{pmatrix}^*=\\
&=\begin{pmatrix}
dd^*&-dc^*&-db^*&da^* \\
-cd^*&cc^*&cb^*&-ca^* \\
-bd^*&bc^*&bb^*&-ba^* \\
ad^*&-ac^*&-ab^*&aa^*
\end{pmatrix}^*=\begin{pmatrix}
dd^*&-cd^*&-bd^*&ad^* \\
-dc^*&cc^*&bc^*&-ac^* \\
-db^*&cb^*&bb^*&-ab^* \\
da^*&-ca^*&-ba^*&aa^*
\end{pmatrix}
\end{split}
\end{equation}

and then we get $R$ equal to

\begin{equation}
\begin{split}
&R^2=\rho\tilde{\rho}=\begin{pmatrix}
aa^*&ab^*&ac^*&ad^* \\
ba^*&bb^*&bc^*&bd^* \\
ca^*&cb^*&cc^*&cd^* \\
da^*&db^*&dc^*&dd^*
\end{pmatrix}\begin{pmatrix}
dd^*&-cd^*&-bd^*&ad^* \\
-dc^*&cc^*&bc^*&-ac^* \\
-db^*&cb^*&bb^*&-ab^* \\
da^*&-ca^*&-ba^*&aa^*
\end{pmatrix}=\\
&=2\begin{pmatrix}
\abs{a}^2\abs{d}^2-ab^*c^*d&-\abs{a}^2cd^*+ab^*\abs{c}^2&-\abs{a}^2bd^*+a\abs{b}^2c^*&\abs{a}^2ad^*-a^2b^*c^*\\
a^*b\abs{d}^2-\abs{b}^2c^*d&-a^*bcd^*+\abs{b}^2\abs{c}^2&-a^*b^2d^*+\abs{b}^2bc^*&\abs{a}^2bd^*-a\abs{b}^2c^*\\
a^*c\abs{d}^2-b^*\abs{c}^2d&-a^*c^2d^*+b^*\abs{c}^2c&-a^*bcd^*+\abs{b}^2\abs{c}^2&\abs{a}^2cd^*-ab^*\abs{c}^2\\
a^*\abs{d}^2d-b^*c^*d^2&-a^*c\abs{d}^2+b^*\abs{c}^2d&-a^*b\abs{d}^2+\abs{b}^2c^*d&\abs{a}^2\abs{d}^2-ab^*c^*d
\end{pmatrix}
\end{split}
\end{equation}

To find the eigenvalues of $R^2$, firstly, it is enough to show that the matrix $R^2$ has three eigenvalues that are equal to zero. To do this it is enough to find by a simple inspection the three eigenvectors $\vec{x}$ such that $R^2\vec{x}=0$. And such three vectors are 

\begin{equation}
\begin{pmatrix}
0\\
b\\
-c\\
0
\end{pmatrix} \,\,\,\,\,\,\,\,\,\,\,\,\,\,\,\,\,\,\, \begin{pmatrix}
b\\
0\\
d\\
0
\end{pmatrix} \,\,\,\,\,\,\,\,\,\,\,\,\,\,\,\,\,\,\, \begin{pmatrix}
0\\
a\\
0\\
c
\end{pmatrix}
\end{equation}

Then, secondly, we need to find trace of $R^2$

\begin{equation}
Tr(R^2)=4(\abs{a}^2\abs{d}^2-ab^*c^*d-a^*bcd^*+\abs{b}^2\abs{c}^2)=4\abs{ad-bc}^2
\end{equation}

which equals to the sum of the eigenvalues of $R^2$. Therefore, the value of the only non-zero eigenvalue of $R^2$ is equal to $4\abs{ad-bc}^2$. And thus the only non-zero eigenvalue of $R$ is $2\abs{ad-bc}$, which then also equals to $C$. 

Now all that is left to complete this subproof is to find the value of $Det(\rho_A)$ and $Det(\rho_B)$:

\begin{equation}
\begin{split}
&Tr_A(\ket{\psi}\bra{\psi})=Tr_B(\ket{\psi}\bra{\psi})=(\abs{a}^2+\abs{b}^2)\ket{0}\bra{1}+(ac^*+bd^*)\ket{0}\bra{1}+\\
&+(ca^*+db^*)\ket{1}\bra{0}+(\abs{c}^2+\abs{d}^2)\ket{1}\bra{1}=\\
&=\begin{pmatrix}
\abs{a}^2+\abs{b}^2&ac^*+bd^* \\
ca^*+db^*&\abs{c}^2+\abs{d}^2\\
\end{pmatrix} \implies Det(\rho_A)=Det(\rho_B)=2\abs{ad-bc}=C
\end{split}
\end{equation}

\hfill $\square$

\end{adjustwidth*}

Now we continue with the proof of the inequality (34). We note that for a pure tripartite state $\ket{\psi}$ of three qubits the value of concurrence $C_{AB}$, that corresponds to the entanglement of formation which measures entanglement between the particles in systems A and B, simplifies. Specifically, the bipartite state $\rho_{AB}$ (density matrix) of composite system with Hilbert space $\mathcal{H}_A\otimes\mathcal{H}_B$ has at most only two non-zero eigenvalues. And therefore $\sqrt{\rho_{AB}\tilde{\rho}_{AB}}$ also has at most two non-zero eigenvalues (both of the statements can be proven by the similar methods that were used in the subproof earlier, but one would have to deal with a general tripartite state which would be even more tedious). Then it follows that $C_{AB}=\lambda_1-\lambda_2$, where $\lambda_1$ and $\lambda_2$ are the two non-zero eigenvalues, and therefore can see that

\begin{equation}
C_{AB}^2=(\lambda_1-\lambda_2)^2=\lambda_1^2+\lambda_2^2-2\lambda_1\lambda_2=Tr(\rho_{AB}\tilde{\rho}_{AB})-2\lambda_1\lambda_2\leq Tr(\rho_{AB}\tilde{\rho}_{AB})
\end{equation}

After getting the same analogous inequality for $C_{AC}^2$ we can sum them together to get 

\begin{equation}
C_{AB}^2+C_{AB}^2\leq Tr(\rho_{AB}\tilde{\rho}_{AB})+Tr(\rho_{AC}\tilde{\rho}_{AC})
\end{equation}

Now we are very close to the resemblance of the inequality (34) that we have been working towards. All that remains for us is to prove that

\begin{equation}
Tr(\rho_{AB}\tilde{\rho_{AB}})+Tr(\rho_{AC}\tilde{\rho_{AC}})=C^2_{A(BC)}
\end{equation} 

To accomplish this let us evaluate $Tr(\rho_{AB}\tilde{\rho_{AB}})$ in terms of the general tripartite pure state $\ket{\psi}$ that we can write like this

\begin{equation}
\ket{\psi}=\sum_{ijk=0}^1 \alpha_{ijk}\ket{ijk}
\end{equation}

So, we start off with the calculation of $\rho_{AB}$:

\begin{equation}
\begin{split}
\rho_{AB}&=Tr_c(\ket{\psi}\bra{\psi})=Tr_c\left(\sum_{ijkmnl}{\alpha^*_{mnl} \alpha_{ijk} \ket{ijk}\bra{mnl}}\right)=\\
&=Tr_c\left(\sum_{ijkmnl}{\alpha^*_{mnl} \alpha_{ijk} \ket{ij}\bra{mn}\otimes\ket{k}\bra{l}}\right)=\sum_{ijkmnl}{\alpha^*_{mnl} \alpha_{ijk} \ket{ij}\bra{mn}}\delta_{lk}=\\
&=\sum_{ijmnk}{\alpha^*_{mnk} \alpha_{ijk} \ket{ij}\bra{mn}}
\end{split}
\end{equation}

Then to find $\tilde{\rho}_{AB}$ we write $\sigma_y\otimes\sigma_y$ in the bra/ket notation as

\begin{equation}
\sigma_y\otimes\sigma_y=-\ket{00}\bra{11}+\ket{01}\bra{10}+\ket{10}\bra{01}-\ket{11}\bra{00}=\sum_{qrst=0}^1{\mathcal{E}_{qs} \mathcal{E}_{tr} \ket{qr}\bra{st}} 
\end{equation}

where $\mathcal{E}_{qs}$ stands for the two-dimensional Levi-Civita 

\begin{equation}
    \mathcal{E}_{qs}=
    \begin{cases}
      1 & \text{if}\,\,\, (q,s)=(0,1)\\
      -1 & \text{if}\,\,\, (q,s)=(1,0)\\
      0 & \text{if}\,\,\, 1=s
    \end{cases}
\end{equation}

So now we can show that

\begin{equation}
\begin{split}
\rho_{AB}(\sigma_y\otimes\sigma_y)&=\sum_{ijmnk} \sum_{qrst}{\alpha^*_{mnk} \alpha_{ijk} \mathcal{E}_{qs} \mathcal{E}_{tr} \ket{ij}\bra{mn}\ket{qr}\bra{st}}=\\
&=\sum_{ijmnk} \sum_{st}{\alpha^*_{mnk} \alpha_{ijk} \mathcal{E}_{ms} \mathcal{E}_{tn} \ket{ij}\bra{st}}
\end{split}
\end{equation}

Since the tensor product of the y-Pauli matrices does not have complex numbers, we get $(\rho_{AB}(\sigma_y\otimes\sigma_y))^*=\rho_{AB}^*(\sigma_y\otimes\sigma_y)$. We will use this to calculate $\rho_{AB}\tilde{\rho}_{AB}$ below:

\begin{equation}
\begin{split}
&\rho_{AB}\tilde{\rho}_{AB}=\rho_{AB}(\sigma_y\otimes\sigma_y)\rho_{AB}^*(\sigma_y\otimes\sigma_y)=\\
&=\sum_{ijmnksti'j'm'n'k's't'}{a^*_{mnk} a_{ijk} \mathcal{E}_{ms} \mathcal{E}_{tn} \alpha_{m'n'k'} \alpha^*_{i'j'k'} \mathcal{E}_{m's'} \mathcal{E}_{t'n'} \ket{ij}\bra{st}\ket{i'j'}\bra{s't'}}=\\
&=\sum_{ijmnki'j'm'n'k's't'}{\alpha^*_{mnk} \alpha_{ijk} \mathcal{E}_{mi'} \mathcal{E}_{j'n} \alpha_{m'n'k'} \alpha^*_{i'j'k'} \mathcal{E}_{m's'} \mathcal{E}_{t'n'} \ket{ij}\bra{s't'}}
\end{split}
\end{equation}

and therefore 

\begin{equation}
Tr(\rho_{AB}\tilde{\rho}_{AB})=\sum_{ijmnk i'j'm'n'k'} {\alpha^*_{mnk} \alpha_{ijk} \mathcal{E}_{mi'} \mathcal{E}_{j'n} \alpha_{m'n'k'} \alpha^*_{i'j'k'} \mathcal{E}_{m'i} \mathcal{E}_{jn'}}
\end{equation}

Now we will use $\mathcal{E}_{mi'} \mathcal{E}_{m'i}=\delta_{mm'}\delta_{i'i}-\delta_{mi}\delta_{i'm'}$ and $\mathcal{E}_{j'n} \mathcal{E}_{jn'}=\delta_{j'j}\delta_{nn'}-\delta_{j'n'}\delta_{nj}$ to show that 

\begin{equation}
\begin{split}
Tr(\rho_{AB}\tilde{\rho}_{AB})&=\sum_{ijkmn i'j'k'm'n'} {\alpha^*_{mnk} \alpha_{ijk} (\delta_{mm'}\delta_{i'i}-\delta_{mi}\delta_{i'm'}) \alpha_{m'n'k'} \alpha^*_{i'j'k'}}\times\\
&\times{(\delta_{j'j}\delta_{nn'}-\delta_{j'n'}\delta_{nj})}=\\
&=\sum_{ijkmni'j'k'}\alpha^*_{mnk} \alpha_{ijk} \alpha_{mnk'} \alpha^*_{ijk'}-\alpha^*_{ink} \alpha_{ijk} \alpha_{i'nk'} \alpha^*_{i'jk'}-\\
&-\alpha^*_{mjk} \alpha_{ijk} \alpha_{mj'k'} \alpha^*_{ij'k'}+\alpha^*_{ijk} \alpha_{ijk} \alpha_{i'j'k'} \alpha^*_{i'j'k'}=\\
&=2Det(\rho_A)-Tr(\rho_B^2)+Tr(\rho_C^2)
\end{split}
\end{equation}

where the last line is true because:

\begin{equation}
\begin{split}
2Det(\rho_A)&=2Det(\sum_{mijk}\alpha^*_{mjk}\alpha_{ijk}\ket{i}\bra{m})=\sum_{mijkm'i'j'k'}{\alpha^*_{mjk} \alpha_{ijk} \alpha^*_{m'j'k'} \alpha_{i'j'k'} \mathcal{E}_{ii'} \mathcal{E}_{mm'}}=\\
&=\sum_{mijkm'i'j'k'}{\alpha^*_{mjk} \alpha_{ijk} \alpha^*_{m'j'k'} \alpha_{i'j'k'} (\delta_{im}\delta_{i'm'}-\delta_{im'}\delta_{i'm})}=\\
&=\sum_{ijki'j'k'}\alpha^*_{ijk} \alpha_{ijk} \alpha^*_{i'j'k'} \alpha_{i'j'k'}-\alpha^*_{i'jk} \alpha_{ijk} \alpha^*_{ij'k'} \alpha_{i'j'k'} 
\end{split}
\end{equation}

which equals to the third and fourth term of (99) and

\begin{equation}
\begin{split}
-Tr(\rho_B^2)&=-Tr(\sum_{ijkmi'j'k'm'}\alpha_{imk}^*\alpha_{ijk}\alpha_{i'm'k'}^*\alpha_{i'j'k'}\ket{j}\bra{m}\ket{j'}\bra{m'})=\\
&=-\sum_{ijki'j'k'}\alpha_{ij'k}^*\alpha_{ijk}\alpha_{i'jk'}^*\alpha_{i'j'k'}
\end{split}
\end{equation}

which equals to the second term of (99) and

\begin{equation}
\begin{split}
Tr(\rho_C^2)&=Tr(\sum_{ijkmi'j'k'm'}\alpha_{ijm}^*\alpha_{ijk}\alpha_{i'j'm'}^*\alpha_{i'j'k'}\ket{k}\bra{m}\ket{k'}\bra{m'})=\\
&=\sum_{ijki'j'k'}\alpha_{ijk'}^*\alpha_{ijk}\alpha_{i'j'k}^*\alpha_{i'j'k'}
\end{split}
\end{equation}

which equals to the first term of (99). Then given that for any $2\times2$ matrices $\rho$ and $\sigma$ this equality is true 

\begin{equation}
Tr(\rho^2)-Tr(\sigma^2)=Det(\sigma)-Det(\rho)
\end{equation}

we finally get

\begin{equation}
Tr(\rho_{AB}\tilde{\rho}_{AB})=2(Det(\rho_A)+Det(\rho_B)-Det(\rho_C))
\end{equation}

and after making the substitution $B\rightarrow C$ we can get

\begin{equation}
Tr(\rho_{AC}\tilde{\rho}_{AC})=2(Det(\rho_A)+Det(\rho_C)-Det(\rho_B))
\end{equation}

and thus 

\begin{equation}
C^2_{AB}+C^2_{AC}\leq Tr(\rho_{AB}\tilde{\rho}_{AB})+Tr(\rho_{AC}\tilde{\rho}_{AC})=4Det(\rho_A)
\end{equation}

We are almost at the end of the proof. Now to proceed we need to remember that $\rho_{BC}$ has only two eigenvalues. This means that only two dimensions of $\rho_{BC}$ and the two-dimensional state $\rho_A$ are needed to construct the tripartite pure state $\ket{\psi}$. Therefore, the state $\rho_{BC}$ can be thought of as the two-dimensional state of a single particle. This means that we can employ the formula that was derived in the Subproof 1 to find the value of the entanglement of formation that measures the entanglement between the particle in system A and the two particles (considered as a single particle with a two-dimentional state) in the composite system with Hilbert space $\mathcal{H_B}\otimes\mathcal{H_C}$. And the squared concurrence $C^2_{A(BC)}$ corresponding to this entanglement of formation must equal to $4Det(\rho_A)$. This follows from the subproof where we showed that $C=2\sqrt{Det(\rho_A)}$ when the joint state of two qubits is pure. And thus, we achieve the result 

\begin{equation}
C^2_{AB}+C^2_{AC}\leq 4Det(\rho_A)=C^2_{A(BC)}
\end{equation}

\hfill $\square$

\end{proof}

\begin{proof}
The concurrence inequality for mixed states is a lot easier to confirm, since we can use the result of the previous proof. And specifically, we can use the fact that 

\begin{equation}
C^2_{AB}(\rho_{AB}^i)+C^2_{AC}(\rho_{AC}^i)\leq C^2_{A(BC)}(\ket{\psi_i}\bra{\psi_i})
\end{equation}

where $\rho_{AB}^i=Tr_C(\ket{\psi_i}\bra{\psi_i})$ and $\rho_{AC}^i=Tr_B(\ket{\psi_i}\bra{\psi_i})$. Then it follows that 

\begin{equation}
\sum_i p_iC^2_{AB}(\rho_{AB}^i)+\sum_i p_iC^2_{AC}(\rho_{AC}^i)\leq \sum_i p_iC^2_{A(BC)}(\ket{\psi_i}\bra{\psi_i})
\end{equation}

where the tripartite mixed state of three qubits is $\rho=\sum_i p_i\ket{\psi_i}\bra{\psi_i}$ such that \newline $\sum_i p_iC^2_{A(BC)}(\ket{\psi_i}\bra{\psi_i})$ has the smallest possible value. Therefore, we get the partial states of composite systems with Hilbert spaces $\mathcal{H}_A\otimes\mathcal{H}_B$ and $\mathcal{H}_A\otimes\mathcal{H}_C$ respectively as

\begin{equation}
\begin{split}
&\rho_{AB}=Tr_C(\rho)=\sum_i p_iTr_C(\ket{\psi_i}\bra{\psi_i})=\sum_i p_i\rho_{AB}^i\\
&\rho_{AC}=Tr_B(\rho)=\sum_i p_iTr_B(\ket{\psi_i}\bra{\psi_i})=\sum_i p_i\rho_{AC}^i\\
\end{split}
\end{equation}

Finally, since, as was mentioned before, a squared concurrence is a convex function, we get 

\begin{equation}
\begin{split}
C^2_{AB}(\rho_{AB})+C^2_{AC}(\rho_{AC})&\leq\sum_i p_iC^2_{AB}(\rho_{AB}^i)+\sum_i p_iC^2_{AC}(\rho_{AC}^i)\leq\\
&\leq\sum_i p_iC^2_{A(BC)}(\ket{\psi_i}\bra{\psi_i})
\end{split}
\end{equation}

\hfill $\square$

\end{proof}

\begin{proof}
We will be focusing mostly on the proof for the entanglement of formation (the full proof for the relative entropy of entanglement is given in [28,32]). The proof for relative entropy of entanglement follows very similarly, so we will not be paying much attention to it. The first thing we do is make use of the following fact. 

\begin{adjustwidth*}{2em}{0em}
\begin{theorem}
for all $\rho\in\mathbb{C}^d\otimes\mathbb{C}^d$, such that $\rho=Tr_{\mathbb{C}^s}(\ket{\psi}\bra{\psi})$, where $\ket{\psi}\in\mathbb{C}^d\otimes\mathbb{C}^d\otimes\mathbb{C}^s$ we have 

\begin{equation}
P\left(\abs{E_F(\rho)-\log(d)+\frac{1}{2\ln(2)}}\leq t\right)\geq 1-e^{-cd^2t^2/\log^2(d)}
\end{equation}

for any fixed $t>0$, where $s\leq cd^2t^2/log^2(d)$ and $c>0$ is some universal constant. 
\end{theorem}
\end{adjustwidth*}

This formula is incredibly difficult to compute. It requires one to be familiar with the L-Lipschitz [29] functions and with lots of lemmas, for example, the Dvoretzky’s Theorem for Lipschitz functions on the sphere [30]. So, we will skip the proof of Theorem 5, the reader may resort for its proof in [31]. So, we will simply continue with the utilisation of definition. With the inequality (112) we will show that the following is true. 

\begin{adjustwidth*}{2em}{0em}
\begin{theorem}
There exists a state $\rho_{ABC}$, such that

\begin{equation}
E_{A(BC)}(\rho_{ABC}^{(\otimes d)})\leq \log_2(d) \,\,\,\,\, \text{and} \,\,\,\,\, E_{AB}(\rho_{AB}^{(\otimes d)})\sim \log_2(d) \sim E_{AC}(\rho_{AC}^{(\otimes d)})
\end{equation}

as $d\rightarrow \infty$. Here $\rho_{ABC}^{(\otimes d)}$ is a state in a system with Hilbert space $\mathcal{H}_A^{(\otimes d)}\otimes\mathcal{H}_B^{(\otimes d)}\otimes\mathcal{H}_C^{(\otimes d)}=\mathbb{C}^d\otimes\mathbb{C}^d\otimes\mathbb{C}^d$. 
\end{theorem}
\end{adjustwidth*}

It turns out that the Theorem 6 is true for a tripartite state $\rho_{ABC}=Tr_{\mathbb{C}^s}(\ket{\psi}\bra{\psi})$ such that $\ket{\psi}\in \mathbb{C}^d\otimes\mathbb{C}^d\otimes\mathbb{C}^d\otimes\mathbb{C}^s$ where $s\sim log_2(d)$ $\implies$ $sd\leq cd^2t^2/log^2(d)$ with $c>0$ being some some universal constant. This means that $\rho_{AB}=Tr_{\mathbb{C}^d\otimes\mathbb{C}^s}(\ket{\psi}\bra{\psi})$ and $\rho_{AC}=Tr_{\mathbb{C}^d\otimes\mathbb{C}^s}(\ket{\psi}\bra{\psi})$. We have set all of the required conditions in order to use the inequality (112). After making our inputs we get 

\begin{equation}
P\left(\abs{E_{AB}(\rho_{AB}^{(\otimes d)})-\log_2(d)+\frac{1}{2\ln(2)}}\leq t\right)\geq 1
\end{equation}

 which is true for any fixed $t>0$ and where $e^{-cd^2t^2/\log^2(d)}$ is vanished because it tends to zero as $d\rightarrow\infty$. So we will always have inequality 

\begin{equation}
\abs{E_{AB}(\rho_{AB}^{(\otimes d)})-\log_2(d)+\frac{1}{2\ln(2)}}\leq t
\end{equation}

to be true. Now let us choose $t=0$. Since left hand side of the inequality must be non-negative, we get 

\begin{equation}
E_{AB}(\rho_{AB}^{(\otimes d)})=\log_2(d)-\frac{1}{2\ln(2)}\sim \log_2(d)
\end{equation}

Equivalently, one gets the same value of $E_{AB}(\rho_{AB}^{(\otimes d)})$. 

While $E_{A(BC)}(\rho_{ABC}^{(\otimes d)})\leq \log_2(d)$ is true simply because the maximum value of $E_{A(BC)}(\rho_{ABC}^{(\otimes d)})$ is $\log_2(d)$. To see why this is true one needs to note there exists maximum entanglement between the particle(s) in the system A and the particles in the composite system with Hilbert space $\mathcal{H}_B\otimes\mathcal{H}_C$ only when $\rho_{AB}^{(\otimes d)}$ or $\rho_{AC}^{(\otimes d)}$ is a maximally entangled state in a system with Hilbert space $\mathcal{H}_A\otimes\mathcal{H}_B=\mathcal{H}_A\otimes\mathcal{H}_C=\mathbb{C}^d\otimes\mathbb{C}^d$. So the tripartite state is 

\begin{equation}
\rho_{ABC}=\frac{\ket{0}\otimes\ket{0}+...+\ket{d-1}\otimes\ket{d-1}}{\sqrt{d}}\otimes\ket{\psi}_C
\end{equation}

where $\ket{\psi}_C$ is an arbitrary pure state in system C. We discussed earlier, at the beginning of chapter 4, that if there is maximal entanglement between the particles in systems A and B then there can be no entanglement with particle(s) in system C. 

Since the tripartite state is pure, we simply need to compute the Von Neumann entropy of entanglement which measures the entanglement between particles in the system A and the particles in the composite with Hilbert space $\mathcal{H}_B\otimes\mathcal{H}_C$. Indeed, we have  

\begin{equation}
\rho_A=Tr_C(Tr_B(\rho_{ABC}))=Tr_B\left(\frac{\ket{0}\otimes\ket{0}+...+\ket{d-1}\otimes\ket{d-1}}{\sqrt{d}}\right)=\frac{1}{d}\sum_{i=0}^{d-1}\ket{i}\bra{i}
\end{equation}

which indeed brings us the familiar result from the Proof 3 where we had 

\begin{equation}
S_{A(BC)}(\rho_{ABC})=-Tr(\rho_A \log_2(\rho_A))= -\log_2(1/d)=\log_2(d)
\end{equation}

So now that we have proven the Theorem 6, we simply need to observe that it implies that there exists $\rho_{ABC}$ such that

\begin{equation}
\max(E_{AB}(\rho_{AB}),E_{AC}(\rho_{AC}))\geq E_{A(BC)}(\rho_{ABC})
\end{equation}

and since we require $f(E_{AB},E_{AC})>\max(E_{AB},E_{AC})$, we can see that the entanglement of formation is not monogamous in the sense of our new definition, because there does not exist a function $f$ such that $f(E_{AB},E_{AC})\leq E_{A(BC)}$ for $\rho_{ABC}$. 

\hfill $\square$

\end{proof}

\begin{proof}
Once again, we will only touch the proof concerning the regularised entanglement of formation, since the proof is very similar for the relative entropy of entanglement. To begin, we first mention that the regularised entanglement of formation $E_F^{\infty}(\rho)$ follows the following two properties:

\begin{adjustwidth*}{0em}{2em}
1) For any bipartite state $\rho_{AB}$ in a system with Hilbert space $\mathcal{H}_A\otimes\mathcal{H}_B$ we have $E_F^{\infty}(\rho_{AB})\leq \min(\log_2(d_A),\log_2(d_B))$, where $d_A$ and $d_B$ are the respective dimensions of the partial states $\rho_A$ and $\rho_B$. This is true because one can show (through the identical operations that were done in the Proof 2 with the use of the Schmidt decomposition) that the maximally entangled state in this case is 

\begin{equation}
\ket{\psi}=\frac{\ket{0}\otimes\ket{0}+...+\ket{\min(d_A,d_B)-1}\otimes\ket{\min(d_A,d_B)-1}}{\sqrt{\min(d_A,d_B)}}
\end{equation}

and thus we can see that the maximum value of $E_F^{\infty}(\rho_{AB})$ is the Von Neumann entropy of entanglement of $\ket{\psi}$ which can be easily computed (result from the Proof 3) to be equal to $\min(\log_2(d_A),\log_2(d_B))$. 

2) For any bipartite antisymmetric state $\alpha_{AB}$ (the definition and the properties of these type of states are best introduced in [33], chapter 2) in a composite system with Hilbert space $\mathcal{H}_A\otimes\mathcal{H}_B$, such that $d_A=d_B$, we have $E_F^{\infty}(\alpha_{AB})\geq c/\log_2(d_A)^t$, where $c,t>0$ are universal constants. This follows from a stronger result: $E_F^{\infty}(\alpha_{AB})\geq c$ where $c>0$ is a universal constant (was shown in [34], chapter IV). 

\end{adjustwidth*}

Now we will start proving the main result. Let us first define a regularised entanglement of formation of the joint antisymmetric state $\alpha_{A_0...A_{2^k}}$ as

\begin{equation}
g_k=E_{A_0(A_1...A_{2^k})}(\alpha_{A_0...A_{2^k}})
\end{equation}

which measures entanglement between the particle(s) in the system $A_0$ with Hilbert space $\mathcal{H}_0=\mathbb{C}^d$ and the particle(s) in the composite system with Hilbert space $\mathcal{H}_1\otimes...\otimes\mathcal{H}_{2^k}=(\mathbb{C}^d)^{(\otimes 2^k)}$, where $d=2^n+1$. Then one can show that the following inequality chain is valid, step by step:

\begin{equation}
\frac{c}{n^t}<\frac{c}{\log_2(d)^t}\leq g_0\leq ... \leq g_n \leq \log_2(d)<n
\end{equation}

We have $c/n^t<c/\log_2(d)^t$ because $n=\log_2(2^n)>log(2^n+1)=log(d)$, Then we have $c/\log_2(d)^t\leq g_0$ by the second property above. It is true that $g_k<g_{k+1}$ because, as we have mentioned in subchapter 4.2, entanglement doesn't increase under local operations, and therefore, it doesn't increase under partial tracing. Then $ g_n \leq \log_2(d)$ by the first property above. And finally $\log_2(d)<n$ because, again, $n=\log_2(2^n)>log(2^n+1)=log(d)$. 

This inequality chain implies that there must exist one or an odd number of $k$'s that are in $\{0,...,n-1\}$, such that 

\begin{equation}
\frac{g_k}{g_{k+1}}\geq 1-\frac{\ln(n^{t+1})}{n}
\end{equation}

This must true because otherwise we would have gotten 

\begin{equation}
\frac{g_0}{g_n}=\prod_{i=0}^{n-1}\frac{g_i}{g_{i+1}}< \left( 1-\frac{\ln(n^{t+1}/c)}{n}\right)^n
\end{equation}

But this must not be the case, because by using the fact that $ln(x)>\frac{x-1}{x}$ for $x>0$ one can show that $\left( 1-\frac{\ln(n^{t+1}/c)}{n}\right)^n\leq c/n^{t+1}$, which means that $g_0/g_n\leq c/n^{t+1}$. And this would contradict the inequality chain (123) because $g_0\leq g_nc/n^{t+1}<c/n^t$. 

Thus, there must exist $0\leq k\leq n-1$ such that 

\begin{equation}
\begin{split}
&E_{A_0(A_1...A_{2^k})}(\alpha_{A_0...A_{2^k}})=E_{A_0(A_{2^k}+1...A_{2^k+1})}(\alpha_{A_0...A_{2^k}})\geq \\
&\geq E_{A_0(A_1...A_{2^{k+1}})}(\alpha_{A_0...A_{2^{k+1}}})\left(1-\frac{\ln(n^{t+1})}{n}\right)
\end{split}
\end{equation}

where $\alpha_{X}$ is an antisymmetric state in system $X$. And as $n\rightarrow\infty$ we get $\ln(n^{t+1})/n\rightarrow 0$ for all $t>0$, therefore 

\begin{equation}
\begin{split}
&E_{A_0(A_1...A_{2^k})}(\alpha_{A_0...A_{2^k}})=E_{A_0(A_{2^k}+1...A_{2^k+1})}(\alpha_{A_0...A_{2^k}})\geq E_{A_0(A_1...A_{2^{k+1}})}(\alpha_{A_0...A_{2^{k+1}}})
\end{split}
\end{equation}

which is equivalent to 

\begin{equation}
E_{AB}(\alpha_{AB})=E_{AC}(\alpha_{AC})\geq E_{ABC}(\alpha_{ABC})
\end{equation}

where $A=A_0$ is a system with Hilbert space $\mathbb{C}^d$, B is a composite system made up of systems $A_1,...,A_{2^k}$ and C is a composite system made up of systems $A_{2^k+1},...,A_{2^{k+1}}$. So B and C both have Hilbert spaces $(\mathbb{C}^d)^{(\otimes 2^k)}$. 

And to conclude, just as in the Proof 9, we again see that there exists $\rho_{ABC}$ such that $\max(E_{AB}(\rho_{AB}),E_{AC}(\rho_{AC}))\geq E_{A(BC)}(\rho_{ABC})$. Therefore, there does not exist a function $f$ such that $f(E_{AB},E_{AC})\leq E_{A(BC)}$ for $\rho_{ABC}$, because we require $f(E_{AB},E_{AC})>\max(E_{AB},E_{AC})$. Thus, the regularised entanglement of formation is not monogamous in the sense of the Definition 14 (where it has the inequality (43) instead of (39)).

\hfill $\square$

\end{proof}

\begin{proof}
We once more only show the proof for the entanglement of formation only, as the proof for the regularised relative entropy of entanglement is very similar. We start with a lemma.

\begin{adjustwidth*}{0em}{2em}
\begin{lemma}
For all tripartite states $\rho_{ABC}$ of a composite system with Hilbert space $\mathcal{H}_A\otimes\mathcal{H}_B\otimes\mathcal{H}_C$ we have

\begin{equation}
E^F_{A(BC)}(\rho_{ABC})\geq E^F_{AB}(\rho_{AB})+E^{R,LOCC^{\leftarrow}}_{AC}(\rho_{AC})
\end{equation}

where $E^F$ stands for the entanglement of formation measuring respective entanglements denoted in the inequality by lower indexes, and $E^{R,LOCC^{\leftarrow}}$ stands for the relative entropy of entanglement which is filtered by the local operations and the one-way classical communication ($LOCC^{\leftarrow}$) measurement. Lower indexes describe between which systems the entanglement is measured, as was expressed earlier. 
\end{lemma}
\end{adjustwidth*}

Specifically, $E^{R,LOCC^{\leftarrow}}$ is defined as 

\begin{equation}
E^{R,LOCC^{\leftarrow}}(\rho)=\sup_{\mathcal{M}}\{\inf_{\sigma}\{S(\mathcal{M}(\rho)||\mathcal{M}(\sigma))\}\}
\end{equation}

where $S(\rho||\sigma)=Tr(\rho log_2(\rho)-\rho log_2(\sigma))$ is the relative entropy and $\mathcal{M}$ denote all of the $LOCC^{\leftarrow}$ measurement maps [35]. 

Now let us see why the inequality (129) is true. Suppose $E^F_{A(BC)}(\rho_{ABC})=\sum_i p_i S_{A(BC)}(\ket{\psi^i_{ABC}}\bra{\psi^i_{ABC}})$ where $\rho_{ABC}=\sum_i p_i \ket{\psi_{ABC}^i}\bra{\psi_{ABC}^i}$ and $S$ is the Von Neumann entropy of entanglement. One can show that (deduced in [36], corollary 2)

\begin{equation}
S_{A(BC)}(\ket{\psi_{ABC}}\bra{\psi_{ABC}})=E^F_{AB}(\rho_{AB})+S_{AC}^{LOCC\leftarrow}(\rho_{AC}||\rho_A\otimes\rho_C)
\end{equation}

Then it follows that 

\begin{equation}
\begin{split}
E^F_{A(BC)}(\rho_{ABC})&=\sum_i p_i[E^F_{AB}(\rho_{AB}^i)+S_{AC}^{LOCC\leftarrow}(\rho_{AC}^i||\rho_A^i\otimes\rho_C^i)]\geq \\
&\geq E^F_{AB}(\rho_{AB})+S_{AC}^{LOCC\leftarrow}(\rho_{AC}||\sum_i p_i\rho_A^i\otimes\rho_C^i)
\end{split}
\end{equation}

where the last inequality is true because $\rho_{AB}=Tr_C(\sum_i p_i \ket{\psi_{ABC}^i}\bra{\psi_{ABC}^i})=$ \newline $=\sum_i p_i Tr_C(\ket{\psi_{ABC}^i}\bra{\psi_{ABC}^i})=\sum_i p_i \rho_{AB}^i$, $\rho_{AC}=\sum_i p_i \rho_{AC}^i$, the entanglement of formation is convex and relative entropy is jointly convex (one can show easily that relative entropy is jointly convex with a use of Lieb’s main concavity theorem [37]). From the definition of $E^{R,LOCC^{\leftarrow}}$ it is then obvious that

\begin{equation}
E^{R,LOCC^{\leftarrow}}_{AC}(\rho_{AC})=\min_{\sigma}S_{AC}^{LOCC\leftarrow}(\rho_{AC}||\sigma)\leq S_{AC}^{LOCC\leftarrow}(\rho_{AC}||\sum_i p_i\rho_A^i\otimes\rho_C^i)
\end{equation}

because $\sum_i p_i\rho_A^i\otimes\rho_C^i$ is one of the possible separable states $\sigma$, but not necessarily the one that gives the smallest value of $S_{AC}^{LOCC\leftarrow}$. Thus, we showed the proof of the lemma. Now we will utilise the inequality (129) along with the following three lemmas that have been proven in their respective references. 

\begin{adjustwidth*}{2em}{0em}
1) First we have (from [38], lemma 3) 

\begin{equation}
E^{R,LOCC^{\leftarrow}}_{AC}(\rho_{AC})\geq \frac{1}{2\ln(2)} \abs{\abs{\rho_{AC}-\sigma_{AC}}}_{LOCC^{\leftarrow}}^2
\end{equation}

where $\sigma_{AC}$ represents joint states in a composite system with Hilbert space $\mathcal{H}_A\otimes\mathcal{H}_C$ such that there is no entanglement between the particles in systems A and C, $\abs{\abs{\rho-\sigma}}_{LOCC^{\leftarrow}}^2$ is a distance between $\rho$ and $\sigma$ filtered by one-way LOCC measurement (given in the introduction of [39], similar concept to the equality (130)). 

2) Then we have (from [39], chapter 2)

\begin{equation}
\abs{\abs{\rho_{AC}-\rho'_{AC}}}_{LOCC^{\leftarrow}}\geq \frac{c_0}{\sqrt{d_Ad_C}}\abs{\abs{\rho_{AC}-\rho'_{AC}}}_1
\end{equation}

for all bipartite states $\rho_{AC}$ and $\rho'_{AC}$, where $\abs{\abs{\rho}}_1=Tr(\sqrt{\rho^{\dagger}\rho})$, $c_0>0$ is a universal constant and $d_A$, $d_C$ are the dimensions of the corresponding systems A and C. 

3) And at last, also have (proven in [40], corollary 4)

\begin{equation}
\abs{E^F_{AC}(\rho_{AC})-E^F_{AC}(\rho'_{AC})}\leq C\log_2(\min(d_A,d_C))\abs{\abs{\rho_{AC}-\rho'_{AC}}}_1^{1/2-\eta}
\end{equation}

for all bipartite states $\rho_{AC}$, $\rho'_{AC}$ and any $\eta>0$.
\end{adjustwidth*}

Thus, we have 

\begin{equation}
\begin{split}
E^{R,LOCC^{\leftarrow}}_{AC}(\rho_{AC})&\geq \frac{1}{2\ln(2)} \abs{\abs{\rho_{AC}-\sigma_{AC}}}_{LOCC^{\leftarrow}}^2\geq \frac{c_0^2}{2\ln(2)d_Ad_C}\abs{\abs{\rho_{AC}-\sigma_{AC}}}_1^2\geq \\
&\geq\frac{c_0^2}{2\ln(2)d_Ad_C}\left(\frac{1}{C\log_2(\min(d_A,d_C))}\right)^{\frac{2}{1/2-\eta}} E^F_{AC}(\rho_{AC})^{\frac{2}{1/2-\eta}}=\\
&=\frac{c}{d_Ad_C\log_2(\min(d_A,d_C))^{4+\eta}}E^F_{AC}(\rho_{AC})^{4+\eta}
\end{split}
\end{equation}

where the last inequality is true because $E^F_{AC}(\sigma_{AC})$ is the entanglement of formation of a separable state, therefore, it equals to zero. To finish the proof simply insert this result into the inequality (129), set $\eta=4$ as our choosing and note that we get the same result when we swap $B$ and $C$, which is the reason we end up with the $\max$ function in the end. 

\hfill $\square$

\end{proof}

\begin{proof}
To understand why $E_{AC}(\rho_{AC})=0$ if $E_{AB}(\rho_{AB})=E_{A(BC)}(\rho_{ABC})$ for the inequality with the entanglement of formation given by

\begin{equation}
\begin{split}
E_{A(BC)}(\rho_{ABC})\geq \max(&E_{AB}(\rho_{AB})+\frac{c}{d_Ad_C\log_2(\min(d_A,d_C))^8}E_{AC}(\rho_{AC})^8,\\
&E_{AC}(\rho_{AC})+\frac{c}{d_Ad_B\log_2(\min(d_A,d_B))^8}E_{AB}(\rho_{AB})^8)
\end{split}
\end{equation}

we must look at the two curves (see Figure 11):

\begin{equation}
E_{A(BC)}=E_{AB}+\frac{c}{d_Ad_C\log_2(\min(d_A,d_C))^8}E_{AC}^8
\end{equation}

\begin{equation}
E_{A(BC)}=E_{AC}+\frac{c}{d_Ad_B\log_2(\min(d_A,d_B))^8}E_{AB}^8
\end{equation}

\begin{center}
\begin{figure}[h]
\centering
\includegraphics[scale=0.3]{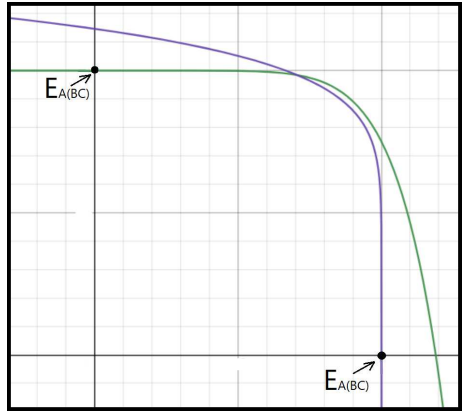}
\caption{An example of plot of the two curves on the space of $(E_{AB},E_{AC)}$.}
\end{figure} 
\end{center}

For the curve (139) we can see that at $E_{AB}=E_{A(BC)}$ we have $E_{AC}=0$ and that it monotonically distances from the line given by $E_{AB}=E_{A(BC)}$ as $E_{AC}$ increases. The same is true for the curve (140), but with B and C swapped. Then one only needs to look at cases where these curves intersect inside the square generated by (48) and the case when they intersect outside. Then it should be obvious why the region generated by (138) doesn't touch the square at any points but $(E_{A(BC)},0)$ and $(0,E_{A(BC)})$. 

\hfill $\square$

\end{proof}

\begin{proof} We will prove the "if and only if" relation from the Theorem 4.

$(\implies)$ Suppose an entanglement measure is monogamous in the sense of the Definition 15. Let us set $x_1=E_{AB}(\rho_{AB})/E_{A(BC)}(\rho_{ABC})$ and $x_1=E_{AC}(\rho_{AC})/E_{A(BC)}(\rho_{ABC})$. Since entanglement doesn't increase under local operations (specifically partial tracing), we have the following three possible outcomes: $x_1> 1$ and $x_2> 1$, $x_1=1$ and $x_2=0$, or $x_1=0$ and $x_2=1$ (last two outcomes follow the Definition 15). Then there exists $\alpha$ such that 

\begin{equation}
x_1^{\alpha}+x_2^{\alpha}\leq 1
\end{equation}

since $x_i^{\alpha}\rightarrow 0$ as $\alpha \rightarrow \infty$, if we have the first outcome. And the inequality (138) is automatically true for the other two outcomes. 

$(\impliedby)$ Suppose there exists $0<\alpha<\infty$ for every fixed dimension $d=dim(\mathcal{H}_A\otimes\mathcal{H}_B\otimes\mathcal{H}_C)$ such that 

\begin{equation}
E_{A(BC)}(\rho_{ABC})\geq (E_{AB}(\rho_{AB})^{\alpha}+E_{AC}(\rho_{AC})^{\alpha})^{1/\alpha}
\end{equation}

for all tripartite states $\rho_{ABC}$ of a composite system with Hilbert space $\mathcal{H}_A\otimes\mathcal{H}_B\otimes\mathcal{H}_C$. Then from the inequality (137) it automatically follows that $E_{A(BC)}(\rho_{ABC})=E_{AB}(\rho_{AB})$ $\implies$ $E_{AC}\rho_{AC})=0$. 

\hfill $\square$

\end{proof}

\begin{proof}
To prove that $(x^{\alpha}+y^{\alpha})^{1/\alpha}\rightarrow \infty$ as $\alpha\rightarrow\infty$ we will first assume that $x>y$. We need to do the following chain of operations (proof derived with the help of an online limit calculator).

\begin{equation}
\begin{split}
\lim_{a\rightarrow\infty}(x^{\alpha}+y^{\alpha})^{1/\alpha}=\lim_{a\rightarrow\infty}e^{\ln((x^{\alpha}+y^{\alpha})^{1/\alpha})}=\lim_{a\rightarrow\infty}e^{\ln(x^{\alpha}+y^{\alpha})/\alpha}=e^{\lim_{a\rightarrow\infty}\ln(x^{\alpha}+y^{\alpha})/\alpha}
\end{split}
\end{equation}

The denominator and the numerator both tend to infinity as $\alpha\rightarrow \infty$, therefore we can use the l'Hopital's rule

\begin{equation}
\begin{split}
&e^{\lim_{a\rightarrow\infty}\frac{d}{da}\ln(x^{\alpha}+y^{\alpha})/(\frac{d}{da}\alpha)}=e^{\lim_{a\rightarrow\infty}(x^{\alpha}\ln(x)+y^{\alpha}\ln(y))/(x^{\alpha}+y^{\alpha})}=\\
&=e^{\lim_{a\rightarrow\infty}(\ln(x)+y^{\alpha}\ln(y)/x^{\alpha})/(1+y^{\alpha}/x^{\alpha})}=e^{\ln(x)}=x
\end{split}
\end{equation}

since for $x>y$ we have $\lim_{\alpha\rightarrow\infty}\left(\frac{y}{x}\right)^{\alpha}=0$. In the case of $y>x$ we need to just swap $x$ and $y$ to see that the result of this limit is $\max(x,y)$. And in the case of $x=y$ the result is trivial. We remind that $x,y\geq 0$ because entanglement measures are lower bounded by zero.  

\hfill $\square$

\end{proof}

\bigskip

\begin{figure}[h]
\centering
\Large
\section{References}
\end{figure}

\normalsize

[1] M. Guta "Introduction to quantum information science" lecture notes, University of Nottingham.

[2] A. Einstein, B. Podolsky, and N. Rosen, “Can quantum-mechanical
description of physical reality be considered complete?”, Phys. Rev.
47, 777 (1935). 

[3] J. S. Bell, “On the Einstein-Podolsky-Rosen paradox.” Physics 1, 195
(1964).

[4] J. F. Clauser, M. A. Horne, A. Shimony, and R. A. Holt, Phys. Rev. Lett. 23, 880 (1980).

[5] M. Plenio and S. Virmani, Quantum Inf. Comput., 7, 1-51 (2007).

[6] J-G. Ren et al., Ground-to-satellite quantum teleportation, Nature 549, 70-73 (2017).

[7] C. H. Bennett, H. J. Bernstein, S. Popescu, and B. Schumacher, Phys. Rev. A 53, 2046 (1996).

[8]  M. Fannes, Commun. Math. Phys. 31, 291 (1973).

[9] V. Vedral, M.B. Plenio, K. Jacobs and P.L. Knight,
Phys. Rev. A 56, 4452 (1997).

[10] M. Christandl and A. Winter, J. Math. Phys 45, 829
(2004). 

[11] P. Hayden, M. Horodecki, and B.M. Terhal, J. Phys. A
34, 6891 (2001).

[12] K. G. H. Vollbrecht, R. F. Werner, Phys. Rev. A 64, no.
6, 062307, (2001).

[13]  R. Alicki and M. Fannes, J.Phys.A:Math.Gen.37 L55 (2004).

[14] P. W. Shor, J. A. Smolin, B. M. Terhal, Phys. Rev. Lett.
86, no. 12, 2681–2684, (2001).

[15] M. B. Hastings, Nature Phys 5, 255–257 (2009).

[16] A. Wehrl, Rev. Mod. Phys. 50, 221, esp. p. 237. (1978).

[17] C.H. Bennett, D.P. DiVincenzo, J.A. Smolin, and W.K.
Wootters, Phys. Rev. A 54, 3824 (1996).

[18] S. Hill and W. K. Wootters, Phys. Rev. Lett. 78, 5022 (1997).

[19] W. K. Wootters, Phys. Rev. Lett. 80, 2245 (1998).

[20] J. Neumann, "Mathematical Foundations of Quantum Mechanics" (1932)

[21] C. Bennett, G. Brassard, C. Crépeau, R. Jozsa, A. Peres and W. Wootters, Phys. Rev. Lett. 70, 1895 (1993).

[22] B. Terhal, IBM J. Res. Dev. 48, 71 (2004).

[23] V. Coffman, J. Kundu, andW. K.Wootters, Phys. Rev. A 61,
052306 (2000).

[24] M.B. Plenio and V. Vedral, Contemp. Phys. 39, 431
(1998).

[25] D. Yang, Phys. Lett. A 360, 249 (2006).

[26] M. Koashi and A. Winter, Phys. Rev. A 69, 022309 (2004).

[27] I. Devetak and A. Winter, Proc. R. Soc. Lond. A 461,
207 (2005).

[28] C. Lancien, S. Di Martino, M. Huber, M. Piani, G. Adesso and A.Winter Phys. Rev. Lett., 117:060501 (2016).

[29] Heinonen J. Lipschitz Functions. In: Lectures on Analysis on Metric Spaces. Universitext. Springer, New York, NY. (2001).

[30] V. D. Milman, Funct. Annal. Appl. 5, 28 (1971).

[31] P. Hayden, D. W. Leung, and A. Winter, Commun. Math. Phys.
265, 95 (2006).

[32] The supplementary material to [28] given at \newline http://link.aps.org/supplemental/10.1103/PhysRevLett.117.060501

[33] I. Jex, G. Alber, S. M. Barnett and A. Delgado Fortschritte der Physik, vol. 51, Issue 2, pp.172-178 (2002).

[34] M. Christandl, N. Schuch, and A. Winter, Commun. Math.
Phys. 311, 397 (2012).

[35] M. Piani, Phys. Rev. Lett. 103, 160504 (2009).

[36] M. Koashi and A. Winter, Phys. Rev. A 69, 022309 (2004).

[37] E. H. Lieb. Adv. Math., 11:267–288 (1973).

[38] F. G. S. L. Brand˜ao, M. Christandl, and J. T. Yard, Commun.
Math. Phys. 306, 805 (2011).

[39] W. Matthews, S.Wehner, and A.Winter, Commun. Math. Phys.
291, 813 (2009).

[40] A.Winter, Commun. Math. Phys. p. 1, ISSN 1432-0916. (2016).

[41] G. Gour and G. Yu, Quantum 2, 81 (2018).

[42] G. Gour and G. Yu, Phys. Rev. A 99, 042305 (2019).

[43] M. A. Nielsen, Phys. Rev. A 61, 064301 (2000).

[44] D. Bruss, J. Math. Phys. 43, 4237 (2002).

[45] Y. Luo and Y. Li. Monogamy of $\alpha$th power entanglement measurement in qubit systems. Ann. Phys., 362:511-520 (2015).

[46] A. K. Ekert, Phys. Rev. Lett. 67, 661 (1991).

[47] U. Vazirani and T. Vidick, Phys. Rev. Lett. 113, 140501 (2014).

[48] X. Ma, B. Dakic, William Naylor1, A. Zeilinger, and P. Walther1, Nat. Phys., 7:399 (2011).

[49] K. Rao, H. Katiyar, T. S. Mahesh, A. Sen, U. Sen and A. Kumar, Phys. Rev. A 88, 022312 (2013). 

[50] A. Almheiri, D. Marolf, J. Polchinski and J. Sully, J. High Energ. Phys. 2013, 62 (2013).

[51] S. W. Hawking, Black hole explosions, Nature 248, 30 (1974).

[52] S. Lloyd and J. Preskill, J. High Energy Phys., 08:126 (2014).

[53] L. Susskind, "Black hole complementarity
and the Harlow-Hayden conjecture.", arXiv: High Energy Physics - Theory (2013)

\end{flushleft}

\end{document}